\documentclass[]{aastex631}

\usepackage{newtxtext}
\usepackage{newtxmath}
\usepackage[T1]{fontenc}
\usepackage{graphicx}	
\usepackage{amsmath}	
\usepackage{amssymb}	
\usepackage{amsfonts}
\usepackage{xcolor}
\usepackage{multirow}
\usepackage{subfigure}
\usepackage{float}
\makeatletter
\let\newfloat\newfloat@ltx
\makeatother
\usepackage{algorithm}
\usepackage{algpseudocode}

\shorttitle{\texttt{LADDER} - Deep Learning Cosmic Distances}
\shortauthors{R. Shah, S. Saha, P. Mukherjee, U. Garain, S. Pal}

\begin{document}

\title{\texttt{LADDER}: Revisiting the Cosmic Distance Ladder with Deep Learning Approaches and Exploring its Applications}

\correspondingauthor{Supratik Pal} 
\email{supratik@isical.ac.in}

\author[0000-0001-7682-9219]{Rahul Shah}
\altaffiliation{Equal contribution.}
\affiliation{Physics and Applied Mathematics Unit \\
Indian Statistical Institute \\
203, B.T. Road, Kolkata 700 108, India}

\author[0000-0003-0013-5143]{Soumadeep Saha}
\altaffiliation{Equal contribution.}
\affiliation{Computer Vision and Pattern Recognition Unit \\
Indian Statistical Institute \\
203, B.T. Road, Kolkata 700 108, India}

\author[0000-0002-2701-5654]{Purba Mukherjee}
\altaffiliation{Equal contribution.}
\affiliation{Physics and Applied Mathematics Unit \\
Indian Statistical Institute \\
203, B.T. Road, Kolkata 700 108, India}

\author[0000-0001-7207-5018]{Utpal Garain}
\affiliation{Computer Vision and Pattern Recognition Unit \\
Indian Statistical Institute \\
203, B.T. Road, Kolkata 700 108, India}

\author[0000-0003-4136-329X]{Supratik Pal}
\affiliation{Physics and Applied Mathematics Unit \\
Indian Statistical Institute \\
203, B.T. Road, Kolkata 700 108, India}




\begin{abstract}
We investigate the prospect of reconstructing the ``cosmic distance ladder'' of the Universe using a novel deep learning framework called \texttt{LADDER} - Learning Algorithm for Deep Distance Estimation and Reconstruction. \texttt{LADDER} is trained on the apparent magnitude data from the Pantheon Type Ia supernovae compilation, incorporating the full covariance information among data points, to produce predictions along with corresponding errors. After employing several validation tests with a number of deep learning models, we pick \texttt{LADDER} as the best performing one. We then demonstrate applications of our method in the cosmological context, including serving as a model-independent tool for consistency checks for other datasets like baryon acoustic oscillations, calibration of high-redshift datasets such as gamma ray bursts, and use as a model-independent mock catalog generator for future probes. Our analysis advocates for careful consideration of machine learning techniques applied to cosmological contexts.
\end{abstract}



\section{Introduction}\label{sec:introduction}
Knowledge of accurate distances to astronomical entities at various redshifts is essential for deducing the expansion history of the Universe. Observationally, however, this task is not simple since there does not exist one single standardizable measure of distances at all scales of cosmological interest. Hence one has to resort to a progressive method of calibrating distances, called the ``cosmic distance ladder'' method, using overlapping regions of potentially different standardizable objects as ``rungs of the ladder''. The conventional distance ladder method \citep{Riess:2023egm} starts with direct measures of geometric distance measures and progresses to calibrating Cepheid variables \citep{Freedman:2023zdo} or Tip of the Red Giant Branch (TRGB) stars \citep{Freedman:2020dne}, and finally Type Ia supernovae (SNIa). Conversely, the ``inverse'' distance ladder begins with cosmology dependent constraints on the sound horizon at drag epoch from the Cosmic Microwave Background (CMB), which is then used to calibrate distances to Baryon Acoustic Oscillations (BAO) and ultimately to SNIa at lower redshifts \citep{Cuesta:2014asa,Camarena:2019rmj}. SNIa are the preferred endpoints for both ladders given their property of being reliable standard candles over a wide redshift range.

A physical theory describing the expansion history of a spatially flat, homogeneous and isotropic universe is given by a cosmological model, which is assumed to be valid over the entire range of observed scales, i.e., from the present epoch ($z=0$) to the epoch of recombination ($z_{\text{CMB}}\sim1100$), with the $\Lambda$ Cold Dark Matter ($\Lambda$CDM) model being the current standard, having six free parameters to be fixed by observations. For a Friedmann-Lema\'{i}tre-Robertson-Walker universe, the cosmic distance-duality relation $d_L=(1+z)^2d_A$ enables switching between luminosity distance $d_L$ and angular diameter distance $d_A$ - the two primary measures of distance in cosmology. The luminosity distances are related to this physical model as $d_{L}=\frac{c(1+z)}{H_{0}}\int_{0}^{z}\frac{dx}{E\left(x\right)}$, where, $E(z)=H(z)/H_0$ is the reduced Hubble parameter, and $H_0=H(z=0)$ is the Hubble constant, signifying the rate of the Universe's expansion today. For sufficiently low redshifts, $d_L$ is well approximated by Hubble's Law, $z = H_0 d_L/c$, offering a means to obtain $H_0$ without assuming a cosmological model. However, of late, inconsistencies have arisen in the concordance model, with the most significant being the tension in the measurement of the Hubble constant ($H_0$) \citep{novosyadlyj,hazra,deg2}. This, and other growing issues with $\Lambda$CDM, has prompted the community to turn either to more complicated cosmological models or to cosmological model-independent (henceforth referred to as simply ``model-independent'') approaches, the second route proving more effective with time.

The simplest method involves cosmography \citep{Visser:2004bf}, which being merely a Taylor expansion of the scale factor does not introduce bias towards any particular cosmological model. There is, however, an ambiguity as to the number of terms to consider in such a series. The aforementioned issues in contemporary cosmology, such as the emergence of tensions, arise when subjected to precision data from observations. This necessitates any alternative method of building the distance ladder to maintain, if not improve, the precision of the data being used. Premature truncation of the cosmographic series may induce significant numerical errors at higher redshifts, while considering higher-order terms raises doubts on convergence. Although alternatives to the Taylor series, such as Pad\'e \citep{Wei:2013jya} and Chebyshev \citep{Capozziello:2017nbu}, help overcome convergence issues to some extent, there still is no clear consensus on the exact number of terms to consider to faithfully mimic the underlying cosmology.

This has motivated the community to resort to reverse engineering by employing model-independent methods for reconstructing distances, and estimation of cosmological parameters therefrom. There have been multiple attempts to reconstruct cosmic distances using Gaussian processes (GP) and genetic algorithms (GA), by various authors, both with present and simulated data from future observatories \citep{Keeley:2020aym,Mukherjee:2021kcu,Arjona:2020axn,Li:2023gpp}. Ambiguity over the choice of kernels, the function dictionary, the mean function, and overfitting concerns with overwhelming errors in data-scarce regions have significantly limited the prospects of these approaches \citep{OColgain:2021pyh, Hwang:2022hla}. This has led to an active use of deep learning with artificial neural networks (ANN) \citep{Wang:2019vxv,Escamilla-Rivera:2021vyw,Olvera:2021jlq,Gomez-Vargas:2022bsm,Gomez-Vargas:2021zyl,Giambagli:2023ngt,Dialektopoulos:2023dhb,Dialektopoulos:2023jam,Mukherjee:2024akt,Zhang:2023xgr,Zhang:2023ucf,Xie:2023ydk,Tang:2020nmw,Wang:2020hmn,Mehrabi:2023tld,Liu:2023rrr} in this domain.

As important accuracy is, measuring distances is limited by experimental precision, due to astrophysical uncertainties, foregrounds, peculiar velocity effects and other practical limitations. Precision is also limited by the available number of data points. These are a critical concern as precision tests are essential for scrutinizing the standard, or alternative, cosmological models. Although both these issues are hoped to be improved upon considerably by upcoming observatories, the use of innovative analysis methods \textit{viz.} Machine Learning (ML) techniques on current data could help overcome these challenges. Another common limitation for a majority of previous ML attempts in cosmology is in the correct and stable prediction of errors at relatively higher redshifts, which makes them unsuitable for undertaking precision cosmological tests when it comes to the issue of tensions. With this motivation, we present our approach - \texttt{LADDER} - Learning Algorithm for Deep Distance Estimation and Reconstruction, which has been designed from the ground up keeping the above considerations in mind. Moreover, almost every straightforward technique fails at extrapolation to any redshift beyond the range of available data in a model-independent manner, due to prediction uncertainties playing the dominant role. Being able to extrapolate beyond the range of available data is lucrative since it could allow simulations of intermediate-redshift data, or in the least, serve as some stable augmentation of currently available data to higher redshifts.

In this spirit, we aim to revisit the cosmic distance ladder by presenting this novel deep learning algorithm \texttt{LADDER} which is trained using the Pantheon SNIa dataset \citep{Scolnic2018}, taking into account the corresponding errors and complete covariances in the data. Our algorithm interpolates from the joint distribution of a randomly chosen subset of the dataset to estimate the target variable and errors simultaneously, and elegantly incorporates correlations and the sequential nature of the data. This leads to predictions that are robust to input noise and outliers and helps make precise predictions even in data-sparse regions. In the following sections, we first outline the datasets and the proposed algorithm, followed by performance validation. We then point out a few cosmological applications that can be explored further using our algorithm. In particular, we demonstrate \texttt{LADDER}'s versatility in conducting consistency checks for a similar SNIa dataset, Pantheon+ \citep{panprelease}. Subsequently, we analyze the implications of our model-independent predictions using the BAO dataset, with regard to their alleged dependence on fiducial cosmology \citep{Sherwin:2018wbu}. We then use the \texttt{LADDER} predictions to calibrate the high redshift Gamma Ray Bursts (GRB) dataset to derive constraints on the $\Lambda$CDM and $w$CDM models. Additionally, we discuss the potential of our deep learning network as a model-independent mock data generator for cosmological studies, and some future directions.

\newpage
\section{Training Dataset}\label{sec:data}
In principle, one can utilize \texttt{LADDER} with any dataset. However, in this article, our primary focus is on achieving a model-independent reconstruction of the cosmological distance ladder. To accomplish this, we employ the Pantheon SNIa compilation \citep{Scolnic2018} as our training dataset. The Pantheon dataset offers several cosmological advantages for model-independent reconstruction of the cosmic distance ladder. Firstly, it remains uncalibrated, which guarantees that it is neither plagued by the choice of cosmological models nor biased by any inherent systematics in otherwise model-independent direct distance measurement calibration methods. Moreover, it involves fewer sources of uncertainties, enhancing its robustness. Additionally, its broad redshift range and diverse data samples help ensure unbiased learning. Notably, no \textit{a priori} priors on the calibration parameter ($M_B$, the absolute magnitude of SNIa in the B-band) are imposed on the ``Pantheon'' dataset during training, either from early-time model-dependent CMB constraints or direct late-time distance measurements.

Pantheon features rich data from 1048 spectroscopically confirmed SNIa spanning a broad range of redshifts $0.01\lesssim{}z<2.3$ with a higher sample density at lower redshifts, and notable sparsity with increasing $z$. This dataset comprises observations on direct measurement of the apparent magnitude ($m$) with the statistical uncertainties ($\Delta{}m$) tabulated at different redshifts ($z$). Additionally, there is a $1048\times1048$ matrix $\mathbf{C}_{\text{sys}}$ corresponding to covariances among the data points. This dataset thus allows for a thorough exploration of cosmic distances covering a wide range of redshifts, and is well-suited for model-independent analyses.

Given knowledge of $M_B$, one can find the luminosity distance $d_L(z)$ independent of any cosmological model. This is expressed by the equation,
\begin{equation}\label{eq:mudL}
    \mu(z)=5\log_{10}\frac{d_L(z)}{\text{Mpc}}+25\:,
\end{equation}
where $\mu(z)=m(z)-M_B$ is the distance modulus. 

The observed apparent magnitudes ($\widetilde{m}$) for each SNIa light curve as measured on Earth depends on the heliocentric ($z_{\text{hel}}$) and CMB frame ($z_{\text{cmb}}$) redshifts. In terms of only $z_{\text{cmb}}$ (i.e. in the absence of peculiar velocities) we have,
\begin{equation}
m\left(z_{\mathrm{cmb}}\right)=\widetilde{m}\left(z_{\mathrm{hel}},z_{\mathrm{cmb}}\right)-5\log_{10}\left(\frac{1+z_{\mathrm{hel}}}{1+z_{\mathrm{cmb}}}\right)\:.
\end{equation}
We then also propagate the errors in $z_{\text{cmb}}$ into $m(z_{\text{cmb}})$. This gives us the data in the final form $m(z_{\text{cmb}})$ \textit{vs} $z_{\text{cmb}}$, which we henceforth refer to as simply $m$ \textit{vs} $z$, with corresponding statistical errors $\Delta{}m$ and covariance matrix $\mathbf{C}_{\text{sys}}$. 

Armed with this data, we aim to train a neural network capable of proficiently learning, and extrapolating to higher redshifts, the apparent magnitude dataset independently of an underlying cosmological model.

\section{Methodology}\label{sec:methodology}
\subsection{Formal Problem Description}
Given the Pantheon dataset, $\mathcal{D}=\{(z_i,m_i,\Delta{}m_i)|\forall{}i\in\{1,\ldots{}N\}\}$, $z_i,m_i\in\mathbb{R}$ which is drawn from some \textit{a priori} unknown distribution, and $\mathbf{C}_{\text{sys}}$, we are interested in estimating the distribution of $\mathcal{P}(M=m|z)\;\forall{}z\in\mathbb{R}^+$ with the assumption, $\mathcal{P}(M=m|z)=\mathcal{N}(\mu_{\theta}(z),\sigma_{\theta}(z))$, for some functions $\mu_{\theta},\sigma_{\theta}$ and some parameter $\theta$. In ML parlance this would be restated as - given $(\mathcal{D},\mathbf{C}_{\text{sys}})$ find $f:\mathcal{Z}\rightarrow\mathbb{R}^2$, such that for any new input $z$, we have,
\begin{equation}
    \min_{f\in\mathcal{F}}\mathcal{E}\Big(\mathcal{N}(f(z)_1,f(z)_2)\Big)\:,
\end{equation}
for a certain class of functions $\mathcal{F}$, which could be a deep learning network and $\mathcal{E}$ is a risk functional. This risk functional is typically the empirical risk,
\begin{equation}
    \mathcal{E}^{\text{empirical}}=\sum_{i=1}^N\ell\Big(\mathcal{N}(m_i,\Delta{}m_i),\mathcal{N}(f(z_i)_1,f(z_i)_2)\Big)\:,
\end{equation}
where $\ell$ is a loss-function, usually Kullback-Leibler (KL) divergence since we are measuring the distance between distributions.

Although our goal is to interpolate from the given points, this problem notably differs from standard regression as the samples are not independent. In particular, since $\mathcal{P}(m_1,m_2,\ldots,m_N|z_1,z_2,\ldots,z_N)\neq\prod_{i=1}^N\mathcal{P}(m_i|z_i)$, our typical empirical risk minimization does not work, and we are left dealing with the following intractable empirical risk,
\begin{equation}\label{eq:intractable}
    \mathcal{E}^{\text{empirical}}=\ell\Big(\mathcal{N}((m_1,m_2,\ldots{}m_N),\Sigma_m),\mathcal{N}((f(z_1)_1,f(z_2)_1,\ldots{}f(z_N)_1),\Sigma_f)\Big)\:.
\end{equation}

\subsection{Our Approach - \texttt{LADDER}}\label{subsec:ladder}
Since our data points are not independent, any predictive model we devise would have to depend on the entire dataset. This presents a challenge, as, whenever we have access to any new data, we must re-adjust our predictive model taking into account the correlations between the new and old data. In order to mitigate this issue without ignoring the correlations between the data instances, we assume that at most $K$ many samples from the dataset are correlated with each other, and rewrite the empirical risk as,
\begin{equation}
    \mathcal{E}^{\text{empirical}}\approx\sum_{\substack{\text{all~combinations}\\j_1,\ldots{}j_K}}\ell\Big(\mathcal{N}\big((m_{j_1},\ldots{}m_{j_K}),\Sigma_m^K\big),\mathcal{N}\big((f(z_{j_1})_1,\ldots{}f(z_{j_K})_1),\Sigma_f^K\big)\Big)\:,
\end{equation}
where $\Sigma_f^K,\Sigma_m^K$ are the predicted and observed covariances respectively, and $f$ is our predictive model. This way, although we are not considering the whole covariance matrix $\Sigma_m$ for each sample, all possible correlations are accounted for, since we are minimizing risk over all observed data-points in aggregate. This motivates our choice of function to be of the form,
\begin{equation}
	\begin{split}
		&f_{\theta}:\mathbb{R}^{2\cdot{}K-1}\rightarrow\mathbb{R}\times\mathbb{R}^+\:,\\
		&f_{\theta}(z;z_{j_1},z_{j_2},\ldots z_{j_{K-1}},m_{j_1},m_{j_2},\ldots{}m_{j_{K-1}})\mapsto(\mu_z,\sigma_z)\:,\\
		&\mathcal{N}(\mu_z,\sigma_z)\approx\mathcal{P}(m|z;z_{j_1},z_{j_2},\ldots{}z_{j_{K-1}},m_{j_1},m_{j_2},\ldots{}m_{j_{K-1}})\:.
	\end{split}
	\label{eq:obj}
\end{equation}
Our objective then is to minimize,
\begin{equation}
\begin{split}
	&\mathcal{E}^{\text{empirical}}\approx\sum_{\substack{\text{all~combinations}\\j_1,\ldots{}j_K}}\sum_{k=1}^N\ell\Big(\mathcal{N}(m_k,\Delta{}m_k),\mathcal{N}(f_{\theta}(z_k;z_{j_1},z_{j_2},\ldots{}m_{j_{K-1}})\Big)\:,\\
    &\ell(P,Q)=D_{KL}(P||Q)=\sum_{x\in\mathcal{X}}P(x)\log\frac{P(x)}{Q(x)}\quad\text{(KL~divergence)}\:.
\end{split}
\end{equation}
$D_{KL}$ is a pseudo-metric measuring the ``distance'' between the distributions $P$ and $Q$. The parameter $\theta$ can be found with an algorithm like stochastic gradient descent.

During training, we first choose $K$ points from $\mathcal{D}$ and designate $K-1$ as ``support'' points, and the remaining point is dubbed the ``query point''. We create a training instance by sampling from $\mathcal{N}(m_{j_2},m_{j_3},\ldots{}m_{j_{K}},\Sigma_m^K)$ to get $\hat{m}_{j_2},\hat{m}_{j_3},\ldots\hat{m}_{j_{K}}$, and create $X=(z_{j_1},z_{j_2},\hat{m}_{j_2},\ldots{}z_{j_{K}},\hat{m}_{j_{K}})$ and $Y=(m_{j_1},\Delta{}m_{j_1})$ (we rearrange the indices such that $z_{j_2}\leq{}z_{j_3}\leq\ldots$). Put simply, the training proxy objective asks - \emph{given these} $K-1$ \emph{points from the dataset, predict} $(m, \Delta m)$ \emph{corresponding to my point of interest}. 

Given $j_1,j_2,\ldots{}j_{K}$, we compute $\Sigma_m^K$ as follows,
\begin{equation}
	\begin{split}
		&\Sigma_m=\mathbf{C}_{\text{sys}}+\mathbb{I}_{N\times{}N}\cdot\Big((\Delta{}m_1)^2,(\Delta{}m_2)^2,\ldots(\Delta{}m_N)^2)\Big)\:,\\
		&[\Sigma_m^K]_{\alpha,\beta}=[\Sigma_m]_{j_{\alpha},j_{\beta}}\quad\forall\alpha,\beta\in\{1,\ldots,K-1\}\:.
	\end{split}
	\label{eq:sigma}
\end{equation}

\begin{figure}[!t]
\centering
\includegraphics[width=0.85\textwidth]{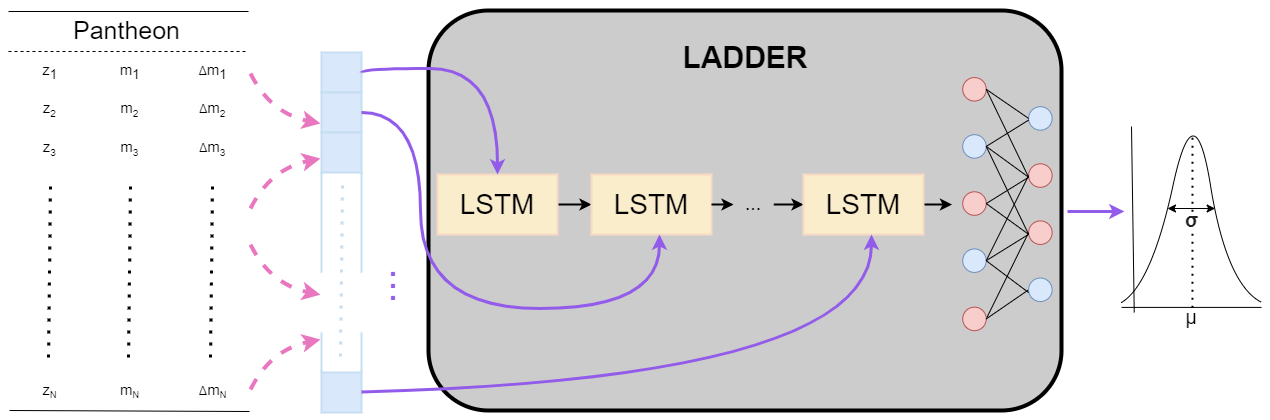}
\caption{Schematic overview of the training algorithm.}
\label{fig:schematic}
\end{figure}

Our job then is to find $\hat{\theta}$, such that $\hat{\theta}=\arg\min_{\theta}\ell[\mathcal{N}(f_{\theta}(X)),\mathcal{N}(m_{j_1},\Delta{}m_{j_1})]$, where $f_{\theta}$ is a suitably chosen deep neural network, with parameters $\theta$. The full algorithm is outlined in Algorithm \ref{algo:training}, with a schematic outlined in Figure~\ref{fig:schematic}.

\begin{algorithm}
\caption{\:\texttt{LADDER} - Learning Algorithm for Deep Distance Reconstruction and Estimation}
\label{algo:training}
    \begin{algorithmic}
    \State Given $\mathcal{D},\mathbf{C}_{\text{sys}}$ and batch size $B$.
    \State Initialize $\theta_0$.
    \While{not StopCondition}
    	\State $l\leftarrow0$
    	\For{$i=1,2,\ldots{}B$}
    		\State $\text{Get }K\text{ samples from }\mathcal{D}$ 
    		\State $\quad\quad\{(z_1,m_1,\Delta{}m_1),\ldots(z_{K},m_{K},\Delta{}m_{K})\}$
    		\State $Y_i=(m_{j_1},\Delta{}m_{j_1})$
    		\State $\hat{m}_{j_2},\hat{m}_{j_3},\ldots\hat{m}_{j_{K}}\sim\mathcal{N}(m_{j_2},m_{j_3},\ldots{}m_{j_{K}},\Sigma_m^K)$ \Comment{(equation~\eqref{eq:sigma})}
    		\State $X_i=(z_{j_1},z_{j_2},\hat{m}_{j_2},\ldots{}z_{j_{K}},\hat{m}_{j_{K}})\quad{}z_{j_2} \leq{}z_{j_3}\ldots$
    		\State $\mu,\sigma=f_{\theta_t}(X_i)_1,f_{\theta_t}(X_i)_2$\Comment{Forward pass.}
    		\State $l\mathrel{+}=D_{KL}\Big(\mathcal{N}(m_{j_1},\Delta{}m_{j_1}),\mathcal{N}(\mu,\sigma)\Big)$
    	\EndFor
    	\State Compute $\nabla_{\theta_t}l\quad\forall\theta_t$
    	\State $\theta_{t+1}=\theta_{t}+\eta\cdot\nabla_{\theta_t}$\Comment{Gradient update (illustrative).}
    	\If{$\ldots$}\Comment{Check if model converged.}
    		\State StopCondition $\leftarrow$ True
    	\EndIf
    \EndWhile
    \end{algorithmic}
\end{algorithm}

Our inference algorithm follows the same basic structure. Given an unseen $z$, we first choose $K-1$ points from $\mathcal{D}$ at random, and sample from $\mathcal{N}((m_{j_2}^{(i)},\ldots,m_{j_{K-1}}^{(i)}),\Sigma_m^K)$ to get $\hat{m}_{j_2}^{(i)},\ldots,\hat{m}_{j_{K-1}}^{(i)}$, and create $X_{\text{unseen}}^{(i)}=(z_{\text{unseen}},z_{j_2}^{(i)},\hat{m}_{j_2}^{(i)},\ldots{}z_{j_{K-1}}^{(i)},\hat{m}_{j_{K-1}}^{(i)})$. We then use $f_{\hat{\theta}}$, to compute $\mu^{(i)},\sigma^{(i)}$. Recall, from equation \eqref{eq:obj}, $\mathcal{N}(\mu_z,\sigma_z)\approx\mathcal{P}(m|z;z_{j_1},z_{j_2},\ldots{}z_{j_{K-1}},m_{j_1},m_{j_2},\ldots{}m_{j_{K-1}})$ and we wish to model $\mathcal{P}(m|z)=\int_{z_1,\ldots{}z_K}\mathcal{P}(m|z;z_{j_1},\ldots{}z_{j_{K-1}},m_{j_1},\ldots{}m_{j_{K-1}})d\mu$. We approximate this with Monte Carlo,
\begin{equation}
	\begin{split}
        \mu_{\text{predicted}}^{(i)},\sigma_{\text{predicted}}^{(i)}=&f_{\hat{\theta}}(X_{\text{unseen}}^{(i)})\:,\\
		\mu=\frac{1}{P}\sum_{i=1}^{P}\mu_{\text{predicted}}^{(i)},&\:\sigma=\frac{1}{P}\sum_{i=1}^{P}\sigma_{\text{predicted}}^{(i)}\:,\\
		\mathcal{P}(m|z)&\approx\mathcal{N}(\mu,\sigma)\:.
	\end{split}
\end{equation}

In addition to being able to model correlations between data points, our approach has another key advantage. Neural networks are universal function approximators \citep{universalApprox} and tend to suffer from overfitting. This problem is exacerbated when dataset sizes are small, and typically we address this with regularization techniques like data augmentation, injecting perturbations or modifying loss functions \citep{augmentation}. Our approach ``augments'' data by sampling from the normal distribution defined by data points and their associated covariances, thus making outputs less sensitive to perturbations of the input. Additionally, since the neural network has to interpolate from a randomly chosen subset of data points, it cannot rely on cues from any single data point, thus making it robust to outliers. Despite random initialization, we found that models trained in this framework reliably converged to the identical final configurations, thus inspiring confidence in this approach's ability to learn underlying causal connections.

\subsection{Model architecture}
We employed two popular neural network architectures for our analysis - the Multi-Layer Perceptron (MLP) \citep{mlp} and Long-Short Term Memory (LSTM) networks \citep{lstm}.

In the multi-layer perceptron model we have $W_h\in\mathbb{R}^{n\times{}d},b_h\in\mathbb{R}^d\:\forall{}h \in\{1,2,\ldots,H\}$, giving,
\begin{equation}
	\begin{split}
		&z_h=\sigma(W_h\cdot{}z_{h-1}+b_h),\\
		&z_0=x \:\:\text{(input)}\:,\\
		&f_{\theta}(x)=z_H\:\:\text{(output)}\:.
	\end{split}
\end{equation}
where $\sigma$ is a non-linear ``transfer'' function like sigmoid, etc. We employed a network with $H=4$ with $W_1\in\mathbb{R}^{2K-1\times16},W_2,W_3\in\mathbb{R}^{16\times16}$ and $W_4\in\mathbb{R}^{16\times2}$.

LSTMs are a type of recurrent neural network that model a sequence of inputs $x_1, x_2, \ldots$ with intermediate representations $h_t$ defined as,
\begin{equation}
	\begin{split}
		&h_t=f_{W_1,\ldots,b_1,\ldots}\Big(x_t,h_{t-1}\Big)\:,\\
		&LSTM(x_1,\ldots,x_T)=h_T\:.
	\end{split}
\end{equation}
We employed a 2-layer LSTM, with a further layer mapping $h_T\rightarrow\mathbb{R}^2$.

We additionally employed batch-normalization and dropouts as deemed appropriate and trained with the AdamW optimizer. We randomly selected 10\% of the training data to serve as a validation set and used grid search to find optimal hyper-parameters. We employed a reduce learning rate on plateau strategy and early stopping using the loss on the validation set.

\subsection{Performance Validation}
To ensure that our proposed algorithm and network architecture are adept at learning the underlying relationship expressed in the data, we first performed some ablation studies. Since we are interested in reconstructing and extrapolating from the data, in addition to accuracy our models must demonstrate some additional cosmological properties. In particular, we expect the cosmological luminosity distance to be monotonic and smooth with respect to redshifts. Thus, we analyze the performance of our models on three metrics \footnote{We use a 80\%-20\% random split of Pantheon for this.} - mean squared error on the validation set, monotonicity as measured by Spearman correlation and smoothness, defined as,
\begin{equation}
	\text{Smoothness}[f_{\theta}]=\frac{1}{n}\sum_{i=1}^{n}|f_{\theta}''(z_i)|\:.
\end{equation}

\begin{figure}[!ht]
\includegraphics[width=\textwidth]{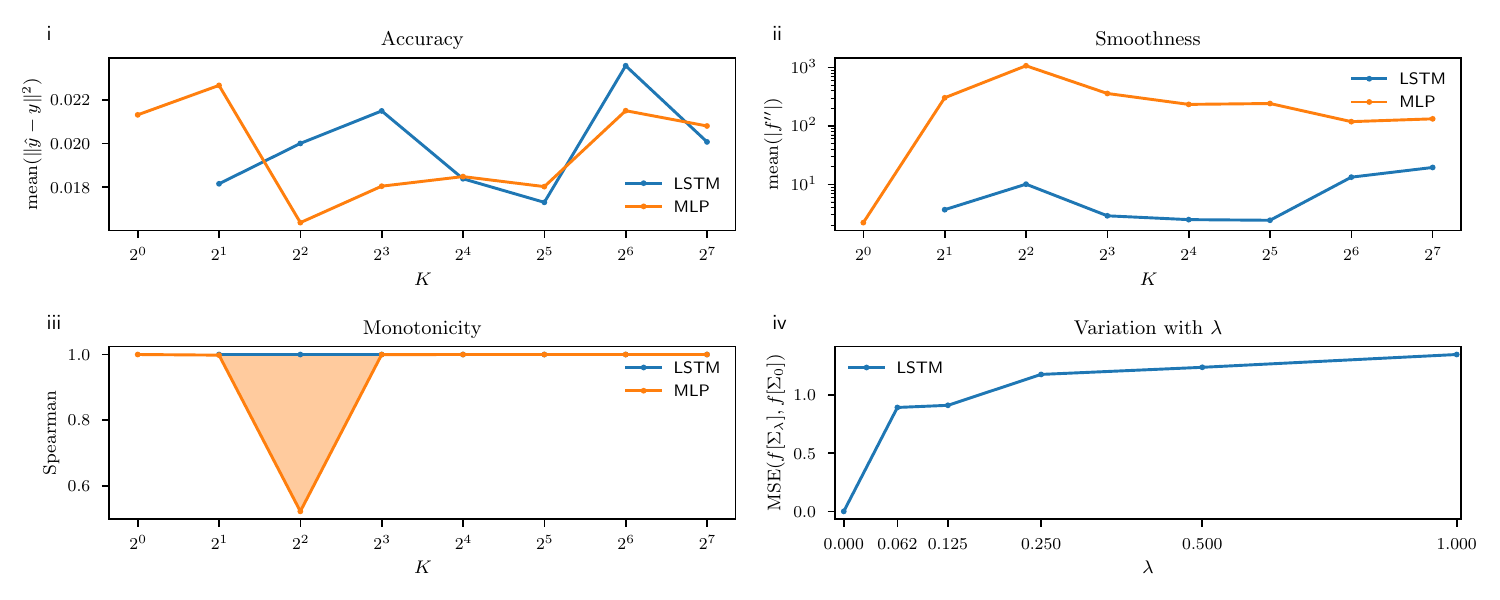}
\caption{Results of ablation studies and validation experiments with various models. (i-iii) Variation of error, smoothness and monotonicity with K respectively. MLP models do not produce smooth, monotonic results (except K=1), and the $K=1$ MLP is outperformed by the LSTM model at roughly the same smoothness and monotonicity. (iv) Prediction variation of models trained on $\Sigma|{\lambda}$ and $\Sigma_0$. When the covariance matrix is progressively corrupted with noise, the predictions change, thus demonstrating our approach's ability to model correlations.}
\label{fig:kDependence}
\end{figure}

We first studied the effect of the parameter $K$ on the predictive performance of our models and found that the best-performing model is the LSTM with $K=32$. In general, the MLP models were found to be lacking in smoothness, and were not reliably monotonic. We also studied how well correlations are captured by our model. To this end, we constructed,
\begin{equation}
	\Sigma_{\lambda}=\lambda\mathbf{N}+(1-\lambda)\mathbf{C}_{\text{sys}}+\mathbb{I}_{N\times{}N}\cdot\Big((\Delta{}m_1)^2,(\Delta{}m_2)^2,\ldots(\Delta{}m_N)^2)\Big)\:,
\end{equation}
where $\mathbf{N}=A\times{}A^T;A_{ij}\sim\mathcal{N}(0,1)$ is a noise matrix, which is by construction symmetric and positive semi-definite. The idea is to ``corrupt'' the covariance matrix and study how the model predictions vary as a result (note that, $\Sigma_{0}=\Sigma_m$). We measured the distance between the distributions ($D_{KL}$) predicted by the model trained with $\Sigma_{\lambda}$ versus those predicted by the model trained with $\Sigma_m$ to see whether the predicted distributions differ. We measured $D_{KL}$, as the noisy covariance matrix is likely to affect the variance predictions as well. We observed that the distance between the predicted distribution of the model trained with $\Sigma_{\lambda}$ and $\Sigma_{0}$ increases steadily with increasing $\lambda$. This shows that the model is capable of picking up cues from the correlations between data points. These results are summarized in Figure~\ref{fig:kDependence}.

\setcounter{table}{0}
\begin{table}[!ht]
	\renewcommand{\thetable}{\arabic{table}}
	\centering
    \caption{Performance of various ML models. $(\downarrow)$ indicates lower is better, $(\uparrow)$ indicates higher is better.}\label{tab:perf}
    \resizebox{1.0\textwidth}{!}{\renewcommand{\arraystretch}{1.0} 
    \setlength{\tabcolsep}{35 pt}
    \begin{tabular}{l r r r r}
        \hline    
            Model                       & MSE $(\downarrow)$    & Monotonicity $(\uparrow)$ & Smoothness $(\downarrow)$ \\    
            \hline    
            \hline    
            kNN(k=5)                    & 0.022116              & 0.99999                   & 90.67500                  \\    
            SVR                         & 0.019358              & $\mathbf{1.0}$            & 3.10633                   \\    
            MLP(K=1)                    & 0.022202              & $\mathbf{1.0}$            & $\mathbf{2.21691}$        \\    
            MLP(K=32)                   & 0.020484              & 0.99997                   & 88.99974                  \\    
            \hline    
            \textbf{\texttt{LADDER}}    & $\mathbf{0.018495}$   & $\mathbf{1.0}$            & 2.30022                   \\ 
            \hline    
    \end{tabular}
    }
\end{table}

We also compare the performance of our model with other regression algorithms like k-Nearest Neighbor Regression (kNNR) and Support Vector Regression (SVR) as measured by accuracy, monotonicity and smoothness (see Table~\ref{tab:perf}). In kNNR, we pick k$(=5)$ nearest instances from the training data and report a weighted sum of their corresponding target values. SVR is a variant of Support Vector Machines, and tries to find a function such that all training instances are within an $\epsilon$-interval of the function. Our algorithm and architecture, outperforms other models in accuracy while maintaining monotonicity and smoothness. Our learning algorithm, outlined in Section \ref{subsec:ladder}, alongside the architecture (LSTM, K=32) is dubbed \texttt{LADDER} - Learning Algorithm for Deep Distance Estimation and Reconstruction. In the subsequent sections, unless explicitly stated, we report results with this approach.

\section{Cosmological Applications of Learning the Distance Ladder}\label{sec:applications}
Given the experimental validation of the performance of our approach in learning the distance ladder, let us now outline a few interesting cosmological applications of the algorithm, which arise as natural consequences of the training process. Of course, our intention is not to do an extensive analysis for each one, which is beyond the scope of the present article anyway. We rather intend to point out possible areas which can be explored further with \texttt{LADDER}, each one of them calling for detailed future investigation.

\subsection{Consistency check for similar datasets in a model-independent approach}

\begin{figure}[!ht]
\gridline{\fig{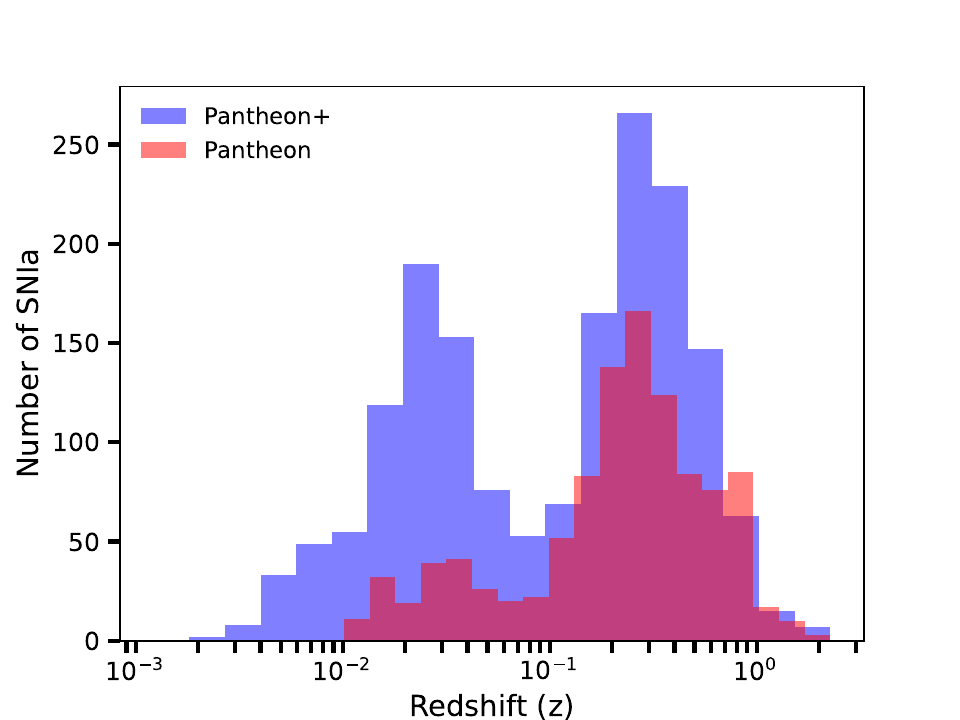}{0.37\textwidth}{Pantheon \textit{vs} Pantheon+ redshift distribution}
          \fig{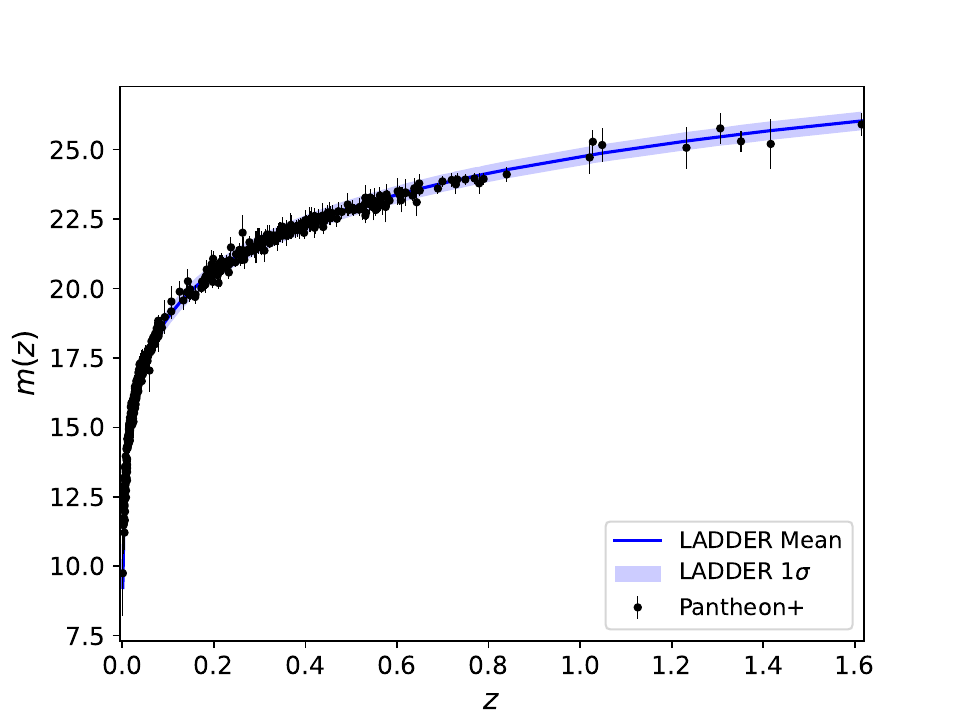}{0.42\textwidth}{Pantheon+ \textit{vs} \texttt{LADDER}}}
\caption{Testing Pantheon+ : Relative redshift distribution of the Pantheon and Pantheon+ datasets (left panel); Comparison between Pantheon-trained \texttt{LADDER} reconstruction \textit{vs} Pantheon+ (right panel).}\label{fig:panplusvalidation}
\end{figure}

Having trained \texttt{LADDER} with Pantheon, we proceed to test the consistency of a comparable SNIa dataset in a model-independent manner. For this, we choose the Pantheon+ compilation \citep{panprelease} which is the latest publicly available SNIa dataset at the time of writing. The major reason to check consistency of Pantheon+ compilation is the following - this compilation comprises data from 1701 light curves representing 1550 distinct SNIa over a redshift range of $0.001\lesssim{}z<2.3$. This range significantly overlaps with that of Pantheon, with a notably high data density at lower redshifts (see the left panel of Figure~\ref{fig:panplusvalidation}). This has potentially significant implications for our understanding of cosmic phenomena, suggesting possible variations in the underlying astrophysical or cosmological processes. This could have a notable impact on the precision of cosmological parameter estimates and the robustness of the underlying physical model, only if the extended dataset with 1701 data points are reliable enough. Several concerns regarding the Pantheon+ dataset have been raised in recent studies. For instance, \citet{Keeley:2022iba} suggest that errors in the covariance matrix may have been overestimated by approximately $5\%$. Additionally, \citet{Perivolaropoulos:2023iqj} have identified evidence of an uncorrected systematic effect, specifically volumetric redshift scatter, in the data. They also highlight a potential physical shift in the absolute magnitude of SNIa at redshift $z=0.005$, assuming a $\Lambda$CDM background. Moreover, \citet{Schoneberg:2022ggi} have reported that Pantheon+ exhibits a preference for a slightly higher value of $\Omega_{m0}$ compared to the original Pantheon dataset within the framework of $\Lambda$CDM cosmology. In the presence of such reports, and the absence of any other SNIa dataset from an independent mission, there is no way to answer directly from data as to how consistent Pantheon+ is. In this context, a well-trained and well-validated ML-based approach may act as a model-independent tool to test for such inconsistencies. Herein lies the role of \texttt{LADDER}.

\begin{figure}[!ht]
\gridline{\fig{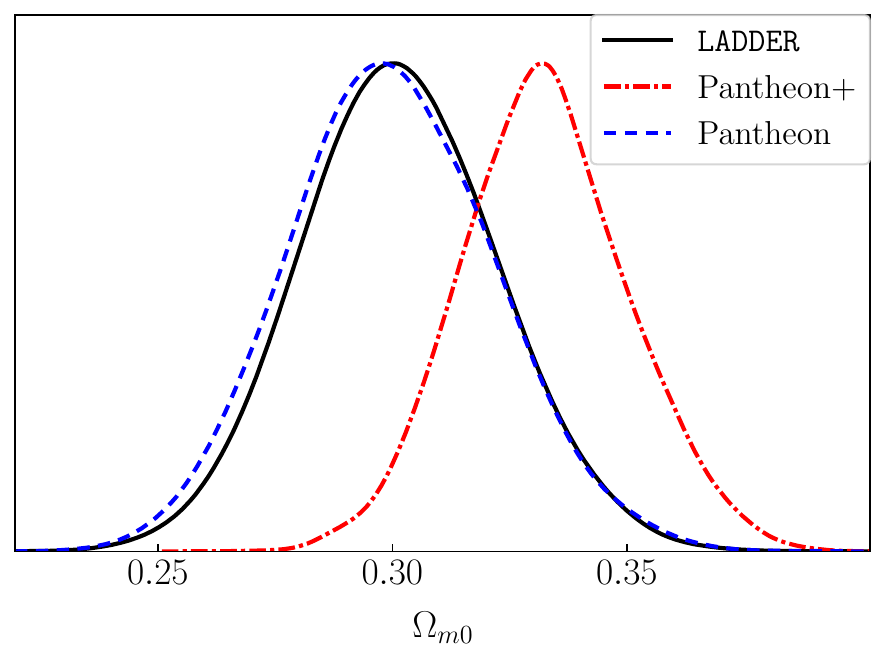}{0.35\textwidth}{$\Lambda$CDM model}\hspace{-2cm}
          \fig{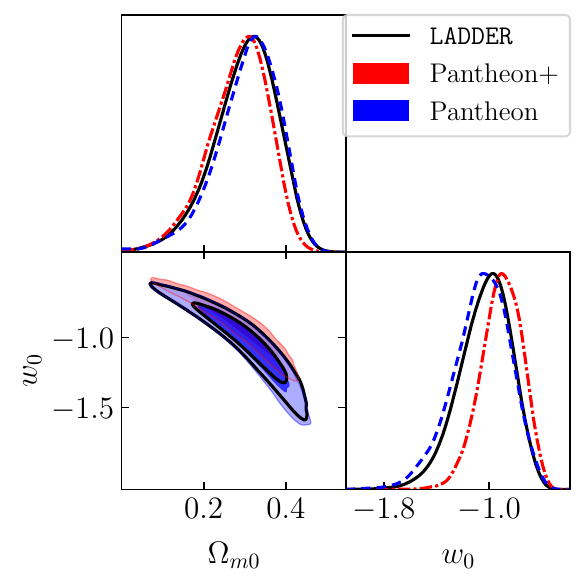}{0.28\textwidth}{$w$CDM model}}
\caption{Cosmological parameter constraints obtained from the Pantheon and Pantheon+ datasets vs \texttt{LADDER} predictions.}
\label{fig:panplusmcmc}
\end{figure}

For an unbiased assessment, we focus solely on the 753 data points in Pantheon+ which are not included in Pantheon. A comparison between these data points and the reconstructions by Pantheon-trained \texttt{LADDER} is presented in the right panel of Figure~\ref{fig:panplusvalidation}. In Figure \ref{fig:panplusmcmc} and Table \ref{tab:panplusconstraints}, we show a comparison of cosmological parameter constraints obtained from the Pantheon and Pantheon+ datasets \textit{vs} \texttt{LADDER} predictions at Pantheon+ redshifts. We see there are negligible mean shifts between Pantheon and \texttt{LADDER}, indicating that \texttt{LADDER} predictions are consistent and in good agreement with Pantheon. Moreover, having trained \texttt{LADDER} on Pantheon, the \texttt{LADDER} predictions at Pantheon+ redshifts show a generic mean shift of the cosmological parameters towards those obtained with Pantheon.

Both Pantheon and \texttt{LADDER} show preference for a phantom equation of state (EoS) $w<-1$. However, we find the constraints on the parameters, $\Omega_{m0}$ and $w_0$, obtained from Pantheon+ have slightly changed in comparison to Pantheon (see \citet{Schoneberg:2022ggi} for previous reports of such trends) or \texttt{LADDER}, where Pantheon+ prefers a higher value of the matter density parameter $\Omega_{m0}$ in the case of the $\Lambda$CDM model. For the $w$CDM model, $\Omega_{m0}$ constraint obtained from Pantheon+ is lower than that from Pantheon or \texttt{LADDER}. Besides, Pantheon+ data shows a preference towards a non-phantom EoS when compared with that of Pantheon or \texttt{LADDER}. This analysis indicates possible inconsistencies in the Pantheon+ compilation, as pointed out earlier. Nonetheless, despite these shifts, the constraints on $\Omega_{m0}$ and $w$ obtained from the three different datasets are found to be consistent with one another at $1\sigma$, for both the models considered. Given that the mean shifts are all within $1\sigma$, shows the consistency of \texttt{LADDER} itself with the two datasets under consideration. However, in order to make any strong comments on these claims one needs to do a full MCMC analysis with \texttt{LADDER} predictions \textit{vis-à-vis} Pantheon+ in combination with CMB, BAO and direct measurement datasets.

\setcounter{table}{1}
\begin{table}[!ht]
	\renewcommand{\thetable}{\arabic{table}}
	\centering
    \caption{Comparison between model parameter constraints obtained using an MCMC analysis with Pantheon and Pantheon+ dataset, vs \texttt{LADDER} predictions.}\label{tab:panplusconstraints}
    \resizebox{1.0\textwidth}{!}{\renewcommand{\arraystretch}{1.5} 
    \setlength{\tabcolsep}{35 pt}
    \begin{tabular}{l c c c}
        \hline
        \multirow{2}{*}{Datasets} & $\Lambda$CDM              & \multicolumn{2}{c}{$w$CDM}                             \\
        \cline{2-4}
                                  & $\Omega_{m0}$            & $\Omega_{m0}$            & $w_0$                        \\ 
        \hline
        \hline
        Pantheon                  & $0.299_{-0.022}^{+0.023}$ & $0.316_{-0.083}^{+0.067}$ & $-1.049_{-0.228}^{+0.199}$ \\

        Pantheon+                 & $0.332_{-0.017}^{+0.018}$ & $0.292_{-0.081}^{+0.065}$ & $-0.902_{-0.162}^{+0.150}$ \\

        \texttt{LADDER}           & $0.301_{-0.021}^{+0.021}$ & $0.308_{-0.083}^{+0.069}$ & $-1.015_{-0.216}^{+0.179}$ \\
        \hline
\end{tabular}
}
\end{table}

Consistency tests, such as these, can be extended to upcoming SNIa datasets, including the recent Dark Energy Survey 5-Year Data Release announcement \citep{DES:2024tys}. This dataset comprises 1829 SNIa in the redshift range of $0.10<z<1.13$, exhibiting a higher data density at higher redshifts compared to Pantheon+. The anticipated release of the actual data holds the potential for an exciting validation study of our ML approach. Furthermore, SNIa datasets from upcoming missions such as Euclid \citep{EuclidTheoryWorkingGroup:2012gxx}, Rubin LSST \citep{lsst2}, Roman Space Telescope \citep{Akeson:2019biv}, the Thirty Meter Telescope (TMT) \citep{TMT}, and the already launched James Webb Space Telescope (JWST) \citep{Gardner_2006} can serve as valuable resources for future efforts towards consistency checks.

\subsection{Pathology test for different datasets in apparent tension}
Since \texttt{LADDER} does not depend on a cosmological model, it can serve as a model-independent test for different types of data, especially those in the same redshift range as the SNIa dataset used for training. Such checks are relevant given the persistent tensions between different datasets. Here, we outline a sketch of the procedure for this with the BAO datasets. In contrast to SNIa measurements, which infer luminosity distances ($d_L$) from apparent magnitude values ($m$) by assuming an absolute magnitude ($M_B$), BAO measures angular distances $d_A/r_d$. The cosmic-distance duality relation, $d_A(z)=(1+z)^{-2}d_L(z)$, connects these quantities, with $r_d$ being the co-moving sound horizon at the drag epoch, which needs to be calibrated for cosmological applications. 

Typically, BAO measurements are combined with CMB data, which can tightly constrain $r_d$ thereby breaking the degeneracy between $r_d$ and $d_A$. In this study, we utilize BAO measurements to establish a connection between SNIa and CMB observations. We calibrate the angular distance from BAO with the \texttt{LADDER} predictions in a model-independent way. This enables the propagation of the CMB constraint on $r_d$ into a constraint on $M_B$, or the SH0ES-prior on $M_B$ into a constraint on $r_d$. We attempt this exercise for two different sets of BAO data:
\begin{itemize}
    \item Transverse Angular BAO ($\theta_{\text{BAO}}$), comprising of 15 model-independent data points in the redshift range $0.11\lesssim{}z\lesssim2.22$, taken from the compilation in \citet{2dbaocompilation}, obtained from SDSS-III LRGs, SDSS-IV blue galaxies and quasars.
    \item Anisotropic BAO ($\alpha_{\text{BAO}}$), consisting of 8 Galaxy+Ly$\alpha$ data points in the redshift range $0.38\lesssim{}z\lesssim2.35$, from the SDSS-III, SDSS-IV DR12 and DR14 data releases, as outlined in \citet{Dialektopoulos:2023dhb}.
\end{itemize}

\begin{figure}[!ht]
\gridline{\hspace{-1cm} \fig{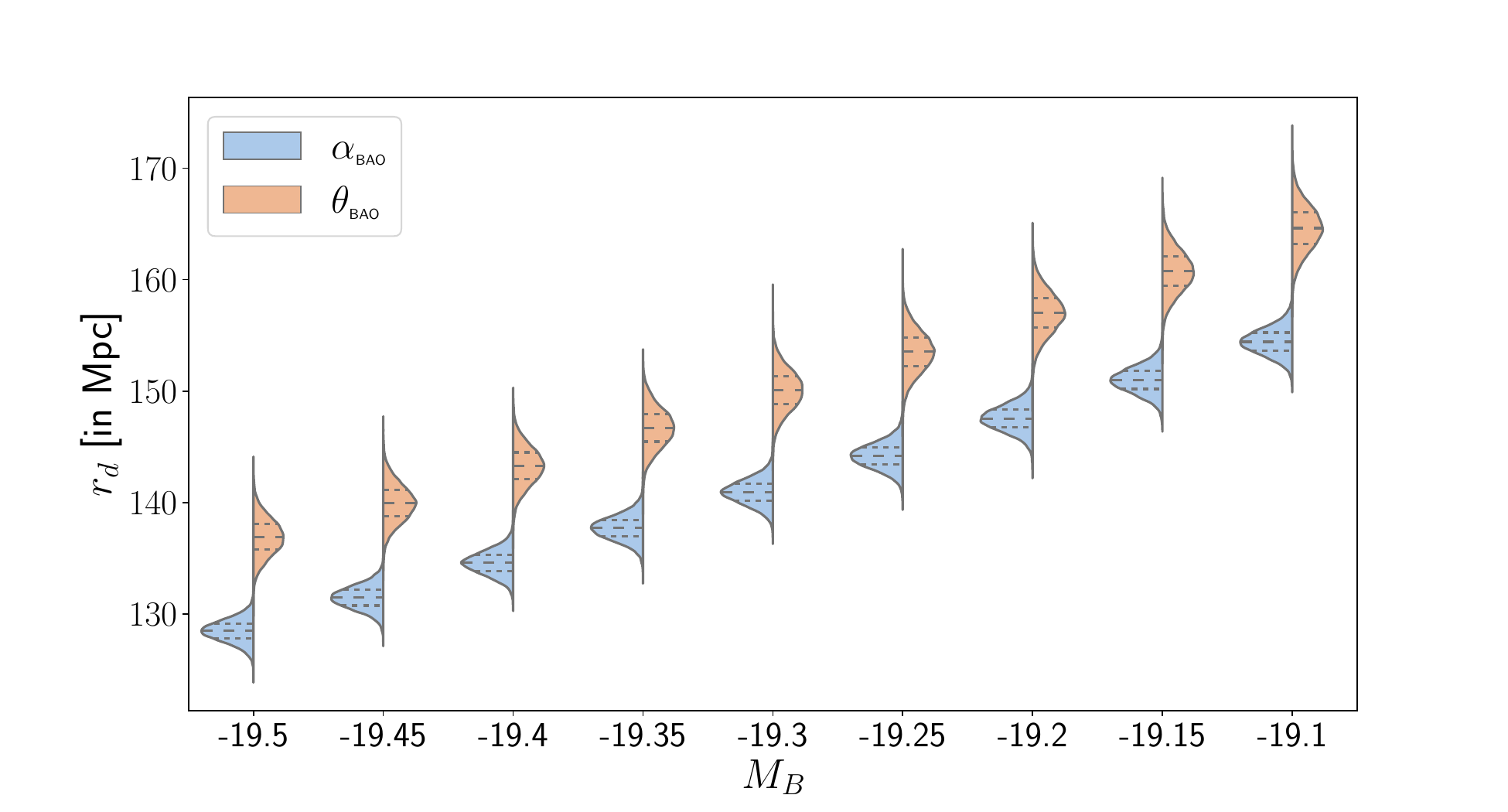}{0.53\textwidth}{Angular \textit{vs} Anisotropic BAO}
          \hspace{-1cm} \fig{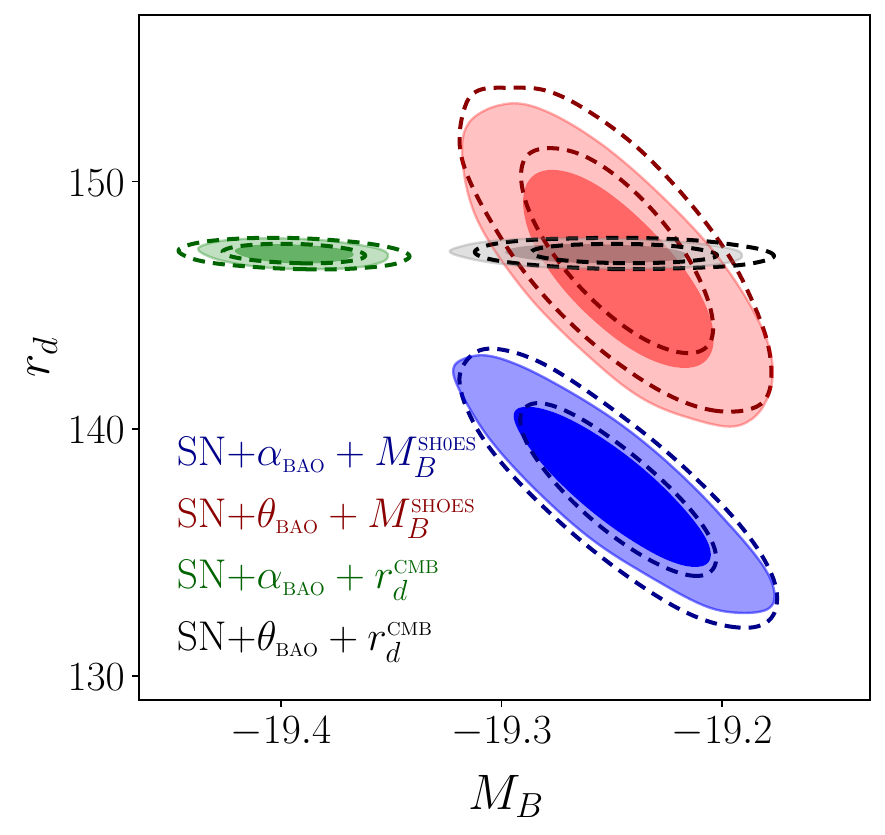}{0.275\textwidth}{\texttt{LADDER} mean predictions \textit{vs} $\Lambda$CDM model}
          \fig{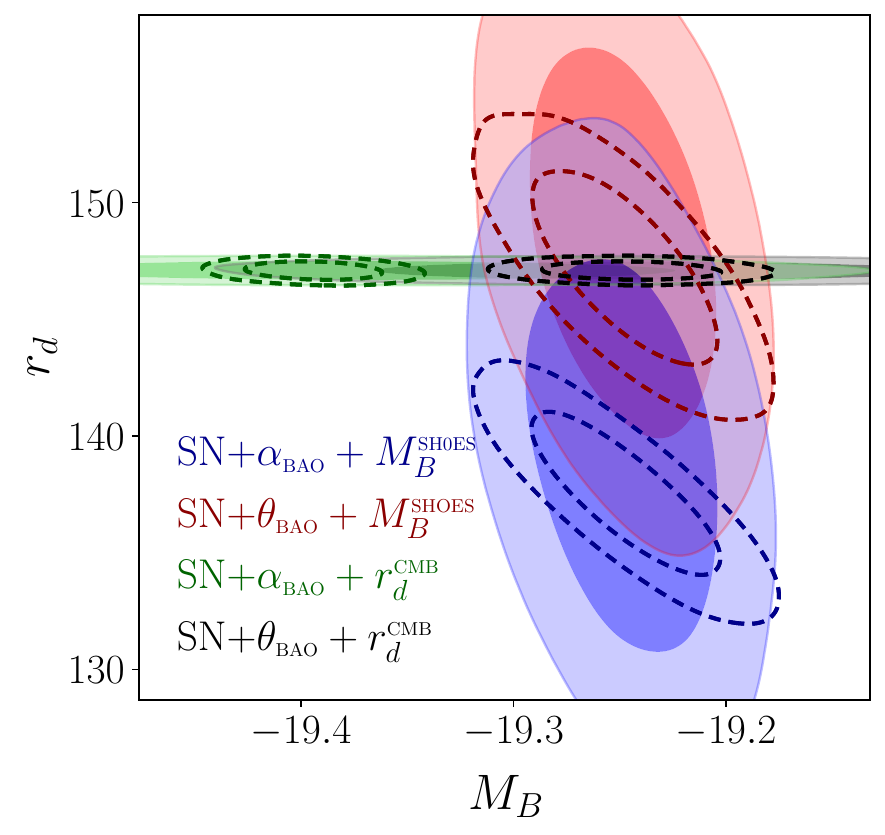}{0.275\textwidth}{\texttt{LADDER} mean + $1\sigma$ uncertainty ~ predictions \textit{vs} $\Lambda$CDM model}}
\caption{Calibration of BAO -- Colored contours are \texttt{LADDER} predictions, dashed ones correspond to benchmark $\Lambda$CDM calibration.}\label{fig:baocalib}
\end{figure}

\setcounter{table}{2}
\begin{table}[!ht]
	\renewcommand{\thetable}{\arabic{table}}
	\centering
    \caption{Comparison between constraints on $M_B$ and $r_d$ [in Mpc], employing \texttt{LADDER} \textit{vs} vanilla $\Lambda$CDM model.}\label{tab:rdmbconstraints}
    \resizebox{1.0\textwidth}{!}{\renewcommand{\arraystretch}{1.5} 
    \setlength{\tabcolsep}{10 pt}
    \begin{tabular}{l c c c c c c}
        \hline
        \multirow{2}{*}{Datasets}                     & \multicolumn{2}{c}{\texttt{LADDER}}                       & \multicolumn{4}{c}{$\Lambda$CDM}                                                                                   \\ 
        \cline{2-7}
                                                      & $M_B$                       & $r_d$                       & $H_0$                      & $\Omega_{m0}$             & $M_B$                       & $r_d$                       \\ 
        \hline
        \hline
        SN+$\alpha_{\text{BAO}}$+$M_B^{\text{SH0ES}}$ & $-19.249^{+0.029}_{-0.029}$ & $137.651^{+2.157}_{-2.153}$ & $73.195_{-1.032}^{+1.054}$ & $0.315_{-0.019}^{+0.019}$ & $-19.248_{-0.029}^{+0.030}$ & $137.484_{-2.228}^{+2.312}$ \\
        SN+$\theta_{\text{BAO}}$+$M_B^{\text{SH0ES}}$ & $-19.248^{+0.028}_{-0.029}$ & $146.423^{+2.697}_{-2.569}$ & $73.372_{-1.010}^{+1.023}$ & $0.302_{-0.021}^{+0.022}$ & $-19.248_{-0.029}^{+0.028}$ & $147.246_{-2.712}^{+2.795}$ \\
        SN+$\alpha_{\text{BAO}}$+$r_d^{\text{CMB}}$   & $-19.394^{+0.018}_{-0.017}$ & $147.090^{+0.250}_{-0.267}$ & $68.437_{-0.826}^{+0.820}$ & $0.314_{-0.019}^{+0.020}$ & $-19.394_{-0.022}^{+0.021}$ & $147.087_{-0.256}^{+0.258}$ \\
        SN+$\theta_{\text{BAO}}$+$r_d^{\text{CMB}}$   & $-19.257^{+0.028}_{-0.027}$ & $147.085^{+0.261}_{-0.253}$ & $73.448_{-1.025}^{+1.042}$ & $0.304_{-0.021}^{+0.022}$ & $-19.245_{-0.027}^{+0.028}$ & $147.088_{-0.256}^{+0.259}$ \\
   \hline
\end{tabular}
}
\end{table}

Initially, we assume different choices of $M_B\in\left[-19.5,-19.1\right]$ and constrain $r_d$ by minimizing the $\chi^2$ metric between the BAO data and the mean \texttt{LADDER} predictions at the corresponding redshifts, taking into account the full covariances associated with the BAO datasets. The direct correlation between the two calibrating parameters is evident from the monotonicity of the constraints depicted in the left-most panel of Figure~\ref{fig:baocalib}.

However, it is noteworthy that discordant constraints on $r_d$ exist between the two BAO datasets for a given value of $M_B$. This discrepancy may imply an internal inconsistency between angular and anisotropic BAO measurements, possibly attributed to the $\Lambda$CDM model dependence in anisotropic BAO data \citep{Carter:2019ulk}, an aspect that calls for further study.

We find out the difference between the vanilla $\Lambda$CDM model-dependent BAO calibration, versus our \texttt{LADDER} based model-independent calibration. In the middle panel of Figure~\ref{fig:baocalib}, we show the joint constraints on the parameters $M_B$ and $r_d$ given different choices of priors - an astrophysical prior on $M_B^{\text{SH0ES}}=-19.2478\pm 0.0294$ \citep{Scolnic2018} and the Planck 2018 CMB prior on $r_d^{\text{CMB}}=147.09\pm 0.26$ Mpc \citep{Pl18VI}, thus breaking the degeneracy between both the parameters. Here SN refers to the Pantheon dataset. The colored contours depict \texttt{LADDER} mean predictions, and the dotted contours are for $\Lambda$CDM. The consistency between the two, albeit with very minor variations, suggests that the true cosmology might resemble something close to the $\Lambda$CDM model. However, the tension between angular \textit{vs} anisotropic BAO measurements is apparent for both methods of calibration here, which is of considerable concern. 

So, we further our pathology test by incorporating the 1$\sigma$ uncertainty predictions from \texttt{LADDER} and redo the above exercise. Notably, the \texttt{LADDER}-based contours widen significantly (see right panel of Figure~\ref{fig:baocalib}). Given uncalibrated Pantheon data, \texttt{LADDER} predicts a minimum as well as a maximum contour spread in comparison to the $\Lambda$CDM model. This data-driven selection of the $M_B$-$r_d$ parameter space can serve as a consistency check for similar exercises with arbitrary models, to be well constrained with Pantheon SN+BAO data. 

It's imperative to recall that values of $M_B$ and $r_d$ significantly influence $H_0$ \citep{Dinda:2022jih,Chen:2024gnu,Mukherjee:2024akt}, hence they need to be chosen with much care in the context of the prevalent Hubble tension. The contour plots in Figure~\ref{fig:baocalib} suggest no apparent tension between Pantheon and $\theta_{\text{BAO}}$ datasets, unlike $\alpha_{\text{BAO}}$. Thus, for any cosmological analysis, the use of angular BAO data, instead of anisotropic BAO, can be recommended as long as late-time datasets are considered in conjunction. Having said that, we should leave a word of caution here. While Table~\ref{tab:rdmbconstraints} suggests an apparent resolution of the $H_0$ tension, an exhaustive Planck CMB analysis is necessary before drawing firm conclusions. Similar efforts can be undertaken with the recent Dark Energy Spectroscopic Instrument (DESI) data release serving as a valuable resource for further consistency checks between BAO data sets \citep{DESI:2024mwx, Cortes:2024lgw, Colgain:2024xqj, Wang:2024rjd}.

\subsection{Model-independent calibration of high-redshift datasets}
Datasets that can potentially yield $d_L$, but lack calibration through well-established anchors in the distance ladder, need to be calibrated by either assuming a cosmological model \citep{Dai:2004tq} or by some model-independent \citep{Liang:2008kx} means. A relevant example involves observations of GRBs which tabulate the GRB spectral peak energy $E^{\text{obs}}_p$ and bolometric fluence $S_{\text{bolo}}$ at different redshifts. 

We investigate if \texttt{LADDER}, trained on uncalibrated Pantheon, can serve as a reliable model-independent calibrator for GRBs. This approach alleviates the challenge of selecting a specific cosmological model for calibration, enabling \texttt{LADDER}-calibrated GRBs as a unique dataset to constrain parameters across diverse cosmological models, avoiding the so-called circularity problem \citep{Ghirlanda:2006ax}.

\begin{figure}[!t]
\gridline{\fig{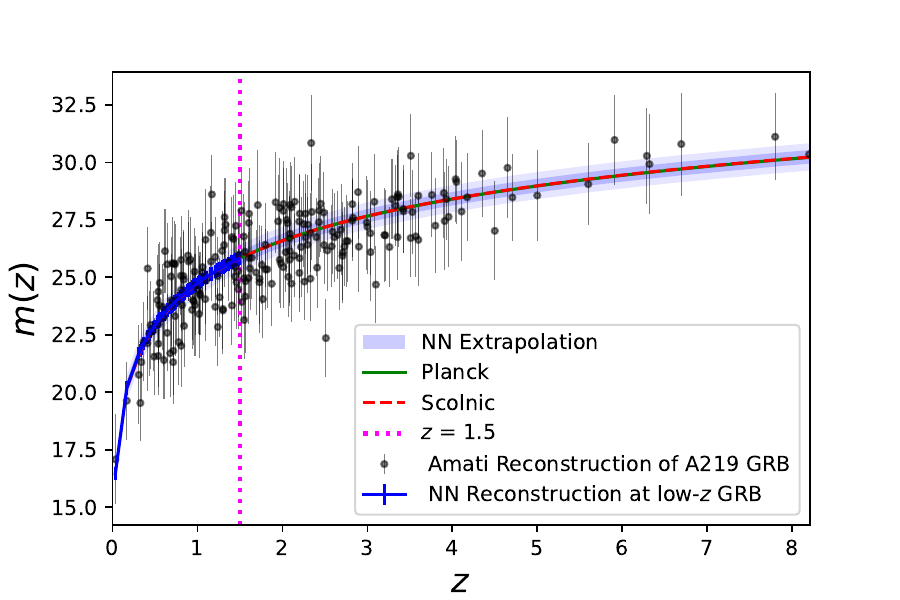}{0.5\textwidth}{\texttt{LADDER} calibrated GRB dataset compared to other methods} \hspace{-2cm} 
          \fig{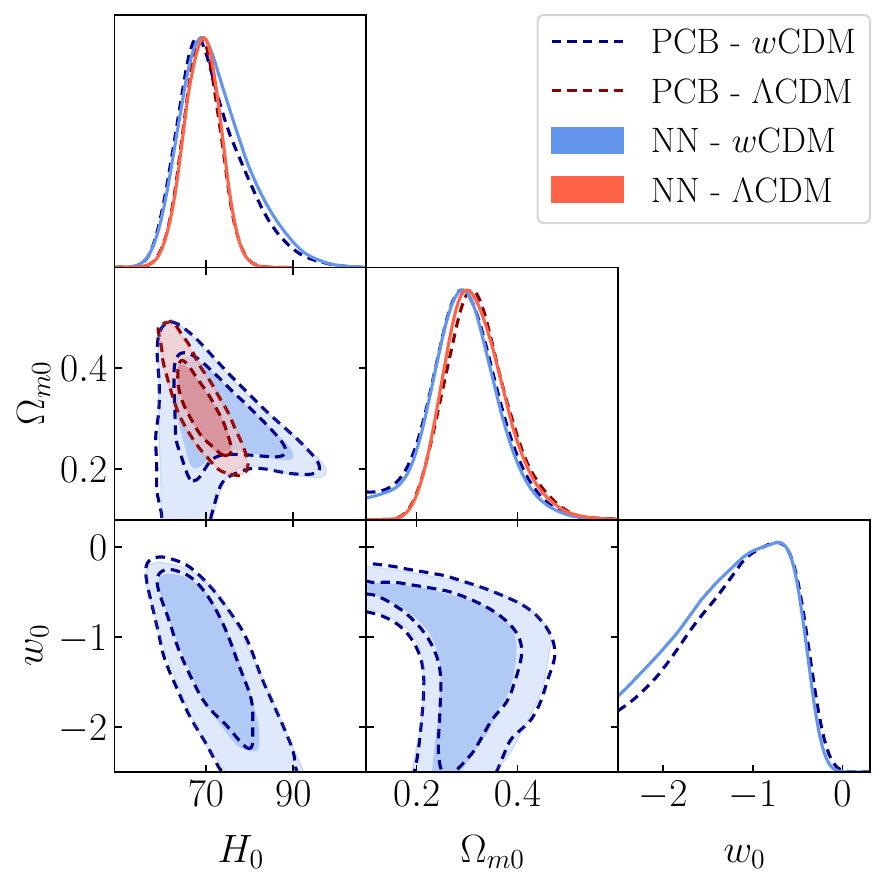}{0.32\textwidth}{Constraints on cosmological model parameters with CC Hubble + calibrated high-$z$ GRB data}}
\caption{GRB calibration output and comparisons.}\label{fig:grbcalib}
\end{figure}

To demonstrate this, we use the GRB A219 sample as outlined in \citet{Liang:2022smf} spanning the redshift range $0.03<z\lesssim8.2$, which correlates the spectral peak energy ($E_p$) and the isotropic equivalent radiated energy ($E_{\text{iso}}$) following Amati relation \citep{Amati:2002ny} as follows, 
\begin{equation}
    \log_{10}\frac{E_{\text{iso}}}{\text{1~erg}}=a+b\,\log_{10}\frac{E_p}{\text{300~keV}}\:.
\end{equation}
Here, $a$ and $b$ are free coefficients, which connects the GRB observables to cosmic distance \citep{Dinda:2022vmb}, such that $E_{\text{iso}}=4\pi{}d^2_L(z)S_{\text{bolo}}(1+z)^{-1}$ and $E_p=E^{\text{obs}}_p(1+z)$. We then split this GRB data into low-redshift ($z<1.5$) and high-redshift ($z>1.5$) samples, consisting of 89 and 130 points respectively. Since \texttt{LADDER} predicts $m$ directly, we express the GRB relation in terms of the same without assuming any prior values of $M_B$ (equation \eqref{eq:mudL}). To this end, the Amati relation is rewritten as, $y^\prime=~a^\prime+b\,x$, where $y^\prime=\log_{10}\left[(1+z)^{-1}\left({S_{\text{bolo}}/{1\text{erg~cm}^{-2}}}\right)\right]+\frac{2}{5}m$. So, $a^\prime=a+2\left(\frac{M_B}{5}+5\right)-\log_{10}\left[4\pi\textrm{(Mpc/cm)}^{2}\right]$ and $b$ are now the free coefficients to be calibrated given $m$ \textit{vs} $z$ \citep{Zhang:2023xgr}. We use the likelihood function by \citet{Reichart_2001} to fit the parameters $a^\prime$ and $b$, using the low-$z$ GRB sample. Table \ref{tab:amati_constraints} shows the constraints obtained on free parameters, $a'$, $b$ and the intrinsic scatter $\sigma_{\text{int}} = \sqrt{{\sigma^2_{y'}}_{\text{int}} + b^2 \, {\sigma^2_{x}}_{\text{int}}}$ of the Amati relation where ${\sigma_{x}}_{\text{int}}$,  ${\sigma_{y'}}_{\text{int}}$ are the intrinsic scatter along the $x$-axis and $y'$-axis respectively. Since \texttt{LADDER} has been trained to learn the distance ladder properly, we then construct a high-$z$ GRB dataset, as shown in the left panel of Figure~\ref{fig:grbcalib}.

\setcounter{table}{3}
\begin{table}[!ht]
	\renewcommand{\thetable}{\arabic{table}}
	\centering
    \caption{Constraints on free parameters of the Amati relation - model \textit{vis-à-vis} \texttt{LADDER} calibration.}\label{tab:amati_constraints}
    \resizebox{1.0\textwidth}{!}{\renewcommand{\arraystretch}{1.7} 
    \setlength{\tabcolsep}{20 pt}
    \begin{tabular}{l c c c c c c}
        \hline
        & $a'$     & $b$  &  ${\sigma_{x}}_{\text{int}}$ &  ${\sigma_{y'}}_{\text{int}}$  &  ${\sigma}_{\text{int}}$ &  $\chi^2_{\text{min}}$ \\
        \hline
        \texttt{LADDER}  &  $9.915_{-0.042}^{+0.041}$ &  $1.599_{-0.087}^{+0.090}$  &  $0.157_{-0.099}^{+0.067}$   &   $0.274_{-0.171}^{+0.083}$  & 0.689 &  43.07  \\
        $\Lambda$CDM     &  $9.932_{-0.041}^{+0.041}$  &  $1.606_{-0.086}^{+0.088}$  &  $0.170_{-0.109}^{+0.056}$  &   $0.259_{-0.169}^{+0.102}$  & 0.711 &  47.61  \\
        $w$CDM           &  $9.924_{-0.041}^{+0.041}$   &  $1.605_{-0.087}^{+0.088}$ &  $0.164_{-0.105}^{+0.058}$ &    $0.258_{-0.169}^{+0.097}$  & 0.699 &  43.21  \\
   \hline
\end{tabular}
}
\end{table}

With this model-independent high-$z$ GRB dataset we obtain constraints on parameter spaces of two cosmological models - baseline $\Lambda$CDM and the wCDM model. For the MCMC analysis, we also incorporate the 32 Cosmic Chronometer (CC) Hubble parameter measurements, as listed in \citet{Dialektopoulos:2023dhb}, along with the full covariance matrix including the systematic and calibration errors \citep{Moresco:2020fbm}, covering the redshift range up to $z\lesssim2$. The results are shown in the right panel of Figure~\ref{fig:grbcalib}. Besides \texttt{LADDER}, for comparison, we further calibrate the Amati coefficients employing both the above-mentioned cosmological models assuming the \textbf{P}antheon SNIa + Planck2018 \textbf{C}MB + \textbf{B}AO (PCB) best-fit model parameter values. The minimized $\chi^2$ of the Amati calibrated low-$z$ GRB data, given in Table \ref{tab:amati_constraints}, indicates that \texttt{LADDER} performs slightly better than $\Lambda$CDM PCB calibration and is comparable to $w$CDM PCB calibration. Following this, we work out the full analysis with the respective model-calibrated datasets. 

\setcounter{table}{4}
\begin{table}[!ht]
	\renewcommand{\thetable}{\arabic{table}}
	\centering
    \caption{Constraints on cosmological models with CC + calibrated high-$z$ GRB datasets.}\label{tab:grbconstraints}
    \resizebox{1.0\textwidth}{!}{\renewcommand{\arraystretch}{1.7} 
    \setlength{\tabcolsep}{30 pt}
    \begin{tabular}{l c c c c}
        \hline
        Model         & \multicolumn{2}{c}{$\Lambda$CDM}                        & \multicolumn{2}{c}{wCDM}                                \\ 
        \cline{2-5}
        Calibrator    & \texttt{LADDER}            & PCB                        & \texttt{LADDER}            & PCB                        \\
        \hline
        \hline
        $H_0$         & $69.203_{-4.150}^{+4.050}$ & $68.996_{-4.095}^{+4.187}$ & $71.224_{-6.365}^{+9.058}$ & $70.265_{-5.915}^{+8.989}$ \\
        $\Omega_{m0}$ & $0.313_{-0.055}^{+0.066}$  & $0.317_{-0.057}^{+0.067}$  & $0.289_{-0.074}^{+0.068}$  & $0.287_{-0.085}^{+0.072}$  \\
        $w_0$         & ...                        & ...                        & $-1.257_{-0.867}^{+0.627}$ & $-1.172_{-0.866}^{+0.599}$ \\
   \hline
\end{tabular}
}
\end{table}

We find that the constraints obtained for $\Lambda$CDM are consistent, irrespective of the calibration method, with only a marginal widening of errors for the model-independent \texttt{LADDER} calibration. Similar conclusions hold when considering a beyond-$\Lambda$CDM model such as $w$CDM. This suggests that regardless of the cosmological model to be constrained, calibration via \texttt{LADDER} can provide a unique high-$z$ GRB dataset in a model-independent setting. Further, Table~\ref{tab:grbconstraints} indicates an apparent trend towards higher $H_0$ values, despite not using any local $H_0$ prior in our analysis. While the associated uncertainties remain considerable, the observed mean shifts in $H_0$ are slightly more pronounced with the implementation of \texttt{LADDER}, compared to the outcomes obtained through model-based calibration. This disparity warrants further investigation and analysis.

GRBs are not the only challenging high-redshift datasets that need calibration independently of any cosmological model. Distance measurements involving high-redshift quasars \citep{Dainotti:2024bth} and early observations from the recently launched JWST \citep{Boylan-Kolchin:2022kae} may also be potential candidates for the application of \texttt{LADDER} calibration. 

\subsection{Other possible applications and future directions}
The persistent tension between $H_0$ inferred from early-time probes, assuming vanilla $\Lambda$CDM, with the directly measured $H_0$ from late-time observations, suggests a potential need for an alternative cosmological model. Various proposals exist in the literature \citep{DiValentino:2021izs,H0Olympics,snowmasstensions,Vagnozzi:2023nrq}, but none fully reconcile CMB-based model-dependent measurements with late-time direct measurements. \texttt{LADDER} offers the opportunity to constrain such alternative models using a broader range of datasets. 

\texttt{LADDER}'s reasonable extrapolation feature to higher redshifts enables generation of synthetic diverse and augmented datasets. This will, in particular, be useful for cosmological forecast studies with upcoming gravitational wave missions, such as LISA \citep{lisa}, ET \citep{Maggiore:2019uih} and DECIGO \citep{DECIGO1}, which often requires generating mock catalogs, inherently relying on a fiducial choice of a cosmological model \citep{Shah:2023,Mukherjee:2023lqr}. We propose that our algorithm could be used as a mock data generator, based solely on the real Pantheon dataset and an assumed value of $M_B$.

Another major cosmological goal is the direct reconstruction of the Hubble parameter. This requires predictions on not just $d_L$, but also its derivatives $d_L^\prime$, with associated errors. GP inherently handle this task by allowing analytical predictions of derivatives and errors, however, they struggle in data-scarce regions and completely fail where no data is available. Neural networks, despite being more performant, do not natively model $d_L^\prime$ and its associated error $\sigma_{d_L^\prime}$. Although attempts to numerically differentiate the network outputs for $d_L$ have been undertaken \citep{Mukherjee:2022yyq,Dialektopoulos:2023dhb,Dialektopoulos:2023jam}, these lead to substantially large uncertainties, rendering final results ineffective when it comes to precision cosmology at higher redshifts. Experiments with \texttt{LADDER} have shown emerging possibilities through potential applications demonstrated in the current study. We plan on expanding upon these results with further detailed analysis and hope to share our findings with the community.

\section{Conclusion}\label{sec:conclusion}
The lack of reliable data points at high redshifts, and the smoothness of the expansion history of our Universe, makes deep learning network models prone to overfitting/underfitting issues, which significantly leads to both inaccurate and imprecise predictions. We hence propose a novel approach in using a deep learning algorithm that learns sequential data while utilizing the full covariance information among data points, applied to cosmological datasets. Our method, part of the new \texttt{LADDER} (Learning Algorithm for Deep Distance Estimation and Reconstruction) suite, is trained on uncalibrated SNIa data from the Pantheon compilation in a manner that is robust to outliers and noise; whilst showing appreciable precision. Our validation experiments show that LSTMs outperform other architectures and ML frameworks in performance and smoothness, perhaps suggesting the importance of capturing the sequential nature of the dataset. The resulting network also exhibits stability, even in data-sparse regions, and enables reliable predictions with associated error bars extending reliably to somewhat beyond the training data range.

Our approach shows promise for contributing to the reconstruction of the cosmic distance ladder in a model-independent manner, as opposed to the conventional cosmological model-dependent approach. We further point out to a few potential applications of \texttt{LADDER}, including the consistency check for similar SNIa datasets, pathology tests for other observational data in apparent tensions, and model-independent calibration of high-redshift measurements. Furthermore, our network could serve as a mock-data generator, capable of reasonable extrapolation without succumbing to overfitting. This capability opens avenues for generating model-independent, Pantheon-based data, at higher redshifts, which is crucial for forecast studies with upcoming cosmological observatories.

While we recognize other approaches, we want to gently emphasize the potential benefits of using advanced learning techniques and attempt to demonstrate how they can optimize information extraction from cosmological data. We hence wish to underscore the innovative nature of employing this new learning algorithm within the cosmological framework and extend our encouragement to the community to consider adopting this algorithm for extracting comprehensive information from cosmological data. We also reiterate that there can be some yet unexplored deep learning algorithm that may perform even better than ours. This work might encourage the community to explore them in different contexts of cosmology. Such efforts would significantly impact the accuracy and reliability of the reconstruction methods and help address yet unresolved issues of cosmology from a different perspective.

\section{Software and third party data repository citations} \label{sec:cite}
The \texttt{LADDER} (Learning Algorithm for Deep Distance Estimation and Reconstruction) suite and data for this article are available on \href{https://github.com/rahulshah1397/LADDER}{GitHub}\footnote{\texttt{LADDER:} \url{https://github.com/rahulshah1397/LADDER}.} under a MIT License and version 1.0 is archived in Zenodo \citep{ladder}.

\software{\href{https://github.com/dfm/emcee}{\textit{emcee}} \citep{Foreman_Mackey_2013}}

\begin{acknowledgments}
We thank the anonymous reviewer for their valuable suggestions towards the improvement of the manuscript. The authors acknowledge the assistance received from the computational facilities made available by the Technology Innovation Hub, ISI Kolkata. RS thanks ISI Kolkata for financial support through Senior Research Fellowship. PM thanks ISI Kolkata for financial support through Research Associateship. UG and SP thank the Department of Science and Technology, Govt. of India for partial support through Grant No. NMICPS/006/MD/2020-21. UG thanks the Indo-French Centre for the Promotion of Advanced Research (IFCPAR/CEFIPRA) for partial support through CSRP Project No. 6702-2.
\end{acknowledgments}

\bibliography{biblio}{}

\begin{thebibliography}{}
\expandafter\ifx\csname natexlab\endcsname\relax\def\natexlab#1{#1}\fi
\providecommand{\url}[1]{\href{#1}{#1}}
\providecommand{\dodoi}[1]{doi:~\href{http://doi.org/#1}{\nolinkurl{#1}}}
\providecommand{\doeprint}[1]{\href{http://ascl.net/#1}{\nolinkurl{http://ascl.net/#1}}}
\providecommand{\doarXiv}[1]{\href{https://arxiv.org/abs/#1}{\nolinkurl{https://arxiv.org/abs/#1}}}

\bibitem[{Abbott {et~al.}(2024)}]{DES:2024tys}
Abbott, T. M.~C., {et~al.} 2024, arXiv.
\newblock \doarXiv{2401.02929}

\bibitem[{Abdalla {et~al.}(2022)}]{snowmasstensions}
Abdalla, E., {et~al.} 2022, JHEAp, 34, 49, \dodoi{10.1016/j.jheap.2022.04.002}

\bibitem[{Adame {et~al.}(2024)}]{DESI:2024mwx}
Adame, A.~G., {et~al.} 2024, arXiv.
\newblock \doarXiv{2404.03002}

\bibitem[{Aghanim {et~al.}(2020)}]{Pl18VI}
Aghanim, N., {et~al.} 2020, Astron. Astrophys., 641, A6,
  \dodoi{10.1051/0004-6361/201833910}

\bibitem[{Akeson {et~al.}(2019)}]{Akeson:2019biv}
Akeson, R., {et~al.} 2019, arXiv:1902.05569, \dodoi{10.48550/arXiv.1902.05569}

\bibitem[{Amati {et~al.}(2002)}]{Amati:2002ny}
Amati, L., {et~al.} 2002, Astron. Astrophys., 390, 81,
  \dodoi{10.1051/0004-6361:20020722}

\bibitem[{Amendola {et~al.}(2013)}]{EuclidTheoryWorkingGroup:2012gxx}
Amendola, L., {et~al.} 2013, Living Rev. Rel., 16, 6,
  \dodoi{10.12942/lrr-2013-6}

\bibitem[{Arjona {et~al.}(2021)Arjona, Lin, Nesseris, \& Tang}]{Arjona:2020axn}
Arjona, R., Lin, H.-N., Nesseris, S., \& Tang, L. 2021, Phys. Rev. D, 103,
  103513, \dodoi{10.1103/PhysRevD.103.103513}

\bibitem[{Bernal {et~al.}(2016)Bernal, Verde, \& Riess}]{deg2}
Bernal, J.~L., Verde, L., \& Riess, A.~G. 2016, JCAP, 10, 019,
  \dodoi{10.1088/1475-7516/2016/10/019}

\bibitem[{Boylan-Kolchin(2023)}]{Boylan-Kolchin:2022kae}
Boylan-Kolchin, M. 2023, Nature Astron., 7, 731,
  \dodoi{10.1038/s41550-023-01937-7}

\bibitem[{Camarena \& Marra(2020)}]{Camarena:2019rmj}
Camarena, D., \& Marra, V. 2020, Mon. Not. Roy. Astron. Soc., 495, 2630,
  \dodoi{10.1093/mnras/staa770}

\bibitem[{Capozziello {et~al.}(2018)Capozziello, D'Agostino, \&
  Luongo}]{Capozziello:2017nbu}
Capozziello, S., D'Agostino, R., \& Luongo, O. 2018, Mon. Not. Roy. Astron.
  Soc., 476, 3924, \dodoi{10.1093/mnras/sty422}

\bibitem[{Carter {et~al.}(2020)Carter, Beutler, Percival, DeRose, Wechsler, \&
  Zhao}]{Carter:2019ulk}
Carter, P., Beutler, F., Percival, W.~J., {et~al.} 2020, Mon. Not. Roy. Astron.
  Soc., 494, 2076, \dodoi{10.1093/mnras/staa761}

\bibitem[{Chen {et~al.}(2024)Chen, Kumar, Ratra, \& Xu}]{Chen:2024gnu}
Chen, Y., Kumar, S., Ratra, B., \& Xu, T. 2024, Astrophys. J. Lett., 964, L4,
  \dodoi{10.3847/2041-8213/ad2e97}

\bibitem[{Colg\'ain {et~al.}(2024)Colg\'ain, Dainotti, Capozziello, Pourojaghi,
  Sheikh-Jabbari, \& Stojkovic}]{Colgain:2024xqj}
Colg\'ain, E.~O., Dainotti, M.~G., Capozziello, S., {et~al.} 2024, arXiv.
\newblock \doarXiv{2404.08633}

\bibitem[{Cort\^es \& Liddle(2024)}]{Cortes:2024lgw}
Cort\^es, M., \& Liddle, A.~R. 2024, arXiv.
\newblock \doarXiv{2404.08056}

\bibitem[{Cuesta {et~al.}(2015)Cuesta, Verde, Riess, \&
  Jimenez}]{Cuesta:2014asa}
Cuesta, A.~J., Verde, L., Riess, A., \& Jimenez, R. 2015, Mon. Not. Roy.
  Astron. Soc., 448, 3463, \dodoi{10.1093/mnras/stv261}

\bibitem[{Dai {et~al.}(2004)Dai, Liang, \& Xu}]{Dai:2004tq}
Dai, Z.~G., Liang, E.~W., \& Xu, D. 2004, Astrophys. J. Lett., 612, L101,
  \dodoi{10.1086/424694}

\bibitem[{Dainotti {et~al.}(2024)Dainotti, Bargiacchi, Lenart, \&
  Capozziello}]{Dainotti:2024bth}
Dainotti, M.~G., Bargiacchi, G., Lenart, A.~L., \& Capozziello, S. 2024, arXiv.
\newblock \doarXiv{2401.11998}

\bibitem[{Di~Valentino {et~al.}(2021)Di~Valentino, Mena, Pan, Visinelli, Yang,
  Melchiorri, Mota, Riess, \& Silk}]{DiValentino:2021izs}
Di~Valentino, E., Mena, O., Pan, S., {et~al.} 2021, Class. Quant. Grav., 38,
  153001, \dodoi{10.1088/1361-6382/ac086d}

\bibitem[{Dialektopoulos {et~al.}(2023)Dialektopoulos, Mukherjee, Levi~Said, \&
  Mifsud}]{Dialektopoulos:2023dhb}
Dialektopoulos, K.~F., Mukherjee, P., Levi~Said, J., \& Mifsud, J. 2023, Eur.
  Phys. J. C, 83, 956, \dodoi{10.1140/epjc/s10052-023-12124-3}

\bibitem[{Dialektopoulos {et~al.}(2024)Dialektopoulos, Mukherjee, Levi~Said, \&
  Mifsud}]{Dialektopoulos:2023jam}
---. 2024, Phys. Dark Univ., 43, 101383, \dodoi{10.1016/j.dark.2023.101383}

\bibitem[{Dinda(2023)}]{Dinda:2022vmb}
Dinda, B.~R. 2023, Int. J. Mod. Phys. D, 32, 2350079,
  \dodoi{10.1142/S0218271823500797}

\bibitem[{Dinda \& Banerjee(2023)}]{Dinda:2022jih}
Dinda, B.~R., \& Banerjee, N. 2023, Phys. Rev. D, 107, 063513,
  \dodoi{10.1103/PhysRevD.107.063513}

\bibitem[{Escamilla-Rivera {et~al.}(2022)Escamilla-Rivera, Carvajal, Zamora, \&
  Hendry}]{Escamilla-Rivera:2021vyw}
Escamilla-Rivera, C., Carvajal, M., Zamora, C., \& Hendry, M. 2022, JCAP, 04,
  016, \dodoi{10.1088/1475-7516/2022/04/016}

\bibitem[{Foreman-Mackey {et~al.}(2013)Foreman-Mackey, Hogg, Lang, \&
  Goodman}]{Foreman_Mackey_2013}
Foreman-Mackey, D., Hogg, D.~W., Lang, D., \& Goodman, J. 2013, Publ. Astron.
  Soc. Pac., 125, 306, \dodoi{10.1086/670067}

\bibitem[{Freedman \& Madore(2023)}]{Freedman:2023zdo}
Freedman, W.~L., \& Madore, B.~F. 2023, arXiv.
\newblock \doarXiv{2308.02474}

\bibitem[{Freedman {et~al.}(2020)Freedman, Madore, Hoyt, Jang, Beaton, Lee,
  Monson, Neeley, \& Rich}]{Freedman:2020dne}
Freedman, W.~L., Madore, B.~F., Hoyt, T., {et~al.} 2020, arXiv,
  \dodoi{10.3847/1538-4357/ab7339}

\bibitem[{Gardner {et~al.}(2006)Gardner, Mather, Clampin, Doyon, Greenhouse,
  Hammel, Hutchings, Jakobsen, Lilly, Long, Lunine, Mccaughrean, Mountain,
  Nella, Rieke, Rieke, Rix, Smith, Sonneborn, Stiavelli, Stockman, Windhorst,
  \& Wright}]{Gardner_2006}
Gardner, J.~P., Mather, J.~C., Clampin, M., {et~al.} 2006, Space Science
  Reviews, 123, 485, \dodoi{10.1007/s11214-006-8315-7}

\bibitem[{Ghirlanda {et~al.}(2006)Ghirlanda, Ghisellini, \&
  Firmani}]{Ghirlanda:2006ax}
Ghirlanda, G., Ghisellini, G., \& Firmani, C. 2006, New J. Phys., 8, 123,
  \dodoi{10.1088/1367-2630/8/7/123}

\bibitem[{Giambagli {et~al.}(2023)Giambagli, Fanelli, Risaliti, \&
  Signorini}]{Giambagli:2023ngt}
Giambagli, L., Fanelli, D., Risaliti, G., \& Signorini, M. 2023, Astron.
  Astrophys., 678, A13, \dodoi{10.1051/0004-6361/202346236}

\bibitem[{G\'omez-Vargas {et~al.}(2023{\natexlab{a}})G\'omez-Vargas, Andrade,
  \& V\'azquez}]{Gomez-Vargas:2022bsm}
G\'omez-Vargas, I., Andrade, J.~B., \& V\'azquez, J.~A. 2023{\natexlab{a}},
  Phys. Rev. D, 107, 043509, \dodoi{10.1103/PhysRevD.107.043509}

\bibitem[{G\'omez-Vargas {et~al.}(2023{\natexlab{b}})G\'omez-Vargas, Esquivel,
  Garc\'\i{}a-Salcedo, \& V\'azquez}]{Gomez-Vargas:2021zyl}
G\'omez-Vargas, I., Esquivel, R.~M., Garc\'\i{}a-Salcedo, R., \& V\'azquez,
  J.~A. 2023{\natexlab{b}}, Eur. Phys. J. C, 83, 304,
  \dodoi{10.1140/epjc/s10052-023-11435-9}

\bibitem[{Hazra {et~al.}(2015)Hazra, Majumdar, Pal, Panda, \& Sen}]{hazra}
Hazra, D.~K., Majumdar, S., Pal, S., Panda, S., \& Sen, A.~A. 2015, Phys. Rev.
  D, 91, 083005, \dodoi{10.1103/PhysRevD.91.083005}

\bibitem[{Hochreiter \& Schmidhuber(1997)}]{lstm}
Hochreiter, S., \& Schmidhuber, J. 1997, Neural Comput., 9, 1735–1780,
  \dodoi{10.1162/neco.1997.9.8.1735}

\bibitem[{Hornik {et~al.}(1989)Hornik, Stinchcombe, \& White}]{universalApprox}
Hornik, K., Stinchcombe, M., \& White, H. 1989, Neural Networks, 2, 359,
  \dodoi{https://doi.org/10.1016/0893-6080(89)90020-8}

\bibitem[{Hwang {et~al.}(2023)Hwang, L'Huillier, Keeley, Jee, \&
  Shafieloo}]{Hwang:2022hla}
Hwang, S.-g., L'Huillier, B., Keeley, R.~E., Jee, M.~J., \& Shafieloo, A. 2023,
  JCAP, 02, 014, \dodoi{10.1088/1475-7516/2023/02/014}

\bibitem[{Keeley {et~al.}(2022)Keeley, Shafieloo, \&
  L'Huillier}]{Keeley:2022iba}
Keeley, R., Shafieloo, A., \& L'Huillier, B. 2022, arXiv.
\newblock \doarXiv{2212.07917}

\bibitem[{Keeley {et~al.}(2021)Keeley, Shafieloo, Zhao, Vazquez, \&
  Koo}]{Keeley:2020aym}
Keeley, R.~E., Shafieloo, A., Zhao, G.-B., Vazquez, J.~A., \& Koo, H. 2021,
  Astron. J., 161, 151, \dodoi{10.3847/1538-3881/abdd2a}

\bibitem[{Li {et~al.}(2024)Li, Keeley, Shafieloo, \& Liao}]{Li:2023gpp}
Li, X., Keeley, R.~E., Shafieloo, A., \& Liao, K. 2024, Astrophys. J., 960,
  103, \dodoi{10.3847/1538-4357/ad0f19}

\bibitem[{Liang {et~al.}(2022)Liang, Li, Xie, \& Wu}]{Liang:2022smf}
Liang, N., Li, Z., Xie, X., \& Wu, P. 2022, Astrophys. J., 941, 84,
  \dodoi{10.3847/1538-4357/aca08a}

\bibitem[{Liang {et~al.}(2008)Liang, Xiao, Liu, \& Zhang}]{Liang:2008kx}
Liang, N., Xiao, W.~K., Liu, Y., \& Zhang, S.~N. 2008, Astrophys. J., 685, 354,
  \dodoi{10.1086/590903}

\bibitem[{Liu {et~al.}(2023)Liu, Hu, Tang, \& Wu}]{Liu:2023rrr}
Liu, L., Hu, L.-J., Tang, L., \& Wu, Y. 2023, Res. Astron. Astrophys., 23,
  125012, \dodoi{10.1088/1674-4527/acf6b3}

\bibitem[{Maggiore {et~al.}(2020)}]{Maggiore:2019uih}
Maggiore, M., {et~al.} 2020, JCAP, 03, 050,
  \dodoi{10.1088/1475-7516/2020/03/050}

\bibitem[{Mandel {et~al.}(2018)Mandel, Sesana, \& Vecchio}]{DECIGO1}
Mandel, I., Sesana, A., \& Vecchio, A. 2018, Class. Quant. Grav., 35, 054004,
  \dodoi{10.1088/1361-6382/aaa7e0}

\bibitem[{Mehrabi(2023)}]{Mehrabi:2023tld}
Mehrabi, A. 2023, arXiv.
\newblock \doarXiv{2301.07369}

\bibitem[{Moresco {et~al.}(2020)Moresco, Jimenez, Verde, Cimatti, \&
  Pozzetti}]{Moresco:2020fbm}
Moresco, M., Jimenez, R., Verde, L., Cimatti, A., \& Pozzetti, L. 2020,
  Astrophys. J., 898, 82, \dodoi{10.3847/1538-4357/ab9eb0}

\bibitem[{Mukherjee {et~al.}(2024{\natexlab{a}})Mukherjee, Dialektopoulos,
  Levi~Said, \& Mifsud}]{Mukherjee:2024akt}
Mukherjee, P., Dialektopoulos, K.~F., Levi~Said, J., \& Mifsud, J.
  2024{\natexlab{a}}, arXiv.
\newblock \doarXiv{2402.10502}

\bibitem[{Mukherjee {et~al.}(2022)Mukherjee, Levi~Said, \&
  Mifsud}]{Mukherjee:2022yyq}
Mukherjee, P., Levi~Said, J., \& Mifsud, J. 2022, JCAP, 12, 029,
  \dodoi{10.1088/1475-7516/2022/12/029}

\bibitem[{Mukherjee \& Mukherjee(2021)}]{Mukherjee:2021kcu}
Mukherjee, P., \& Mukherjee, A. 2021, Mon. Not. Roy. Astron. Soc., 504, 3938,
  \dodoi{10.1093/mnras/stab1054}

\bibitem[{Mukherjee {et~al.}(2024{\natexlab{b}})Mukherjee, Shah, Bhaumik, \&
  Pal}]{Mukherjee:2023lqr}
Mukherjee, P., Shah, R., Bhaumik, A., \& Pal, S. 2024{\natexlab{b}}, Astrophys.
  J., 960, 61, \dodoi{10.3847/1538-4357/ad055f}

\bibitem[{Novosyadlyj {et~al.}(2014)Novosyadlyj, Sergijenko, Durrer, \&
  Pelykh}]{novosyadlyj}
Novosyadlyj, B., Sergijenko, O., Durrer, R., \& Pelykh, V. 2014, JCAP, 05, 030,
  \dodoi{10.1088/1475-7516/2014/05/030}

\bibitem[{Nunes {et~al.}(2020)Nunes, Yadav, Jesus, \&
  Bernui}]{2dbaocompilation}
Nunes, R.~C., Yadav, S.~K., Jesus, J.~F., \& Bernui, A. 2020, Mon. Not. Roy.
  Astron. Soc., 497, 2133, \dodoi{10.1093/mnras/staa2036}

\bibitem[{\'O~Colg\'ain \& Sheikh-Jabbari(2021)}]{OColgain:2021pyh}
\'O~Colg\'ain, E., \& Sheikh-Jabbari, M.~M. 2021, Eur. Phys. J. C, 81, 892,
  \dodoi{10.1140/epjc/s10052-021-09708-2}

\bibitem[{Olvera {et~al.}(2022)Olvera, G\'omez-Vargas, \&
  V\'azquez}]{Olvera:2021jlq}
Olvera, J. d. D.~R., G\'omez-Vargas, I., \& V\'azquez, J.~A. 2022, Universe, 8,
  120, \dodoi{10.3390/universe8020120}

\bibitem[{Perivolaropoulos \& Skara(2023)}]{Perivolaropoulos:2023iqj}
Perivolaropoulos, L., \& Skara, F. 2023, Mon. Not. Roy. Astron. Soc., 520,
  5110, \dodoi{10.1093/mnras/stad451}

\bibitem[{Reichart(2001)}]{Reichart_2001}
Reichart, D.~E. 2001, The Astrophysical Journal, 553, 235,
  \dodoi{10.1086/320630}

\bibitem[{Riess \& Breuval(2023)}]{Riess:2023egm}
Riess, A.~G., \& Breuval, L. 2023, arXiv.
\newblock \doarXiv{2308.10954}

\bibitem[{Sanger \& Baljekar(1958)}]{mlp}
Sanger, T., \& Baljekar, P.~N. 1958, Psychological review, 65 6, 386.
\newblock \url{https://api.semanticscholar.org/CorpusID:12781225}

\bibitem[{Sch\"oneberg {et~al.}(2022{\natexlab{a}})Sch\"oneberg,
  Franco~Abell\'an, P\'erez~S\'anchez, Witte, Poulin, \&
  Lesgourgues}]{H0Olympics}
Sch\"oneberg, N., Franco~Abell\'an, G., P\'erez~S\'anchez, A., {et~al.}
  2022{\natexlab{a}}, Phys. Rept., 984, 1,
  \dodoi{10.1016/j.physrep.2022.07.001}

\bibitem[{Sch\"oneberg {et~al.}(2022{\natexlab{b}})Sch\"oneberg, Verde,
  Gil-Mar\'\i{}n, \& Brieden}]{Schoneberg:2022ggi}
Sch\"oneberg, N., Verde, L., Gil-Mar\'\i{}n, H., \& Brieden, S.
  2022{\natexlab{b}}, JCAP, 11, 039, \dodoi{10.1088/1475-7516/2022/11/039}

\bibitem[{Scolnic {et~al.}(2022)}]{panprelease}
Scolnic, D., {et~al.} 2022, Astrophys. J., 938, 113,
  \dodoi{10.3847/1538-4357/ac8b7a}

\bibitem[{{Scolnic} {et~al.}(2018){Scolnic}, {Jones}, {Rest}, {Pan},
  {Chornock}, {Foley}, {Huber}, {Kessler}, {Narayan}, {Riess}, {Rodney},
  {Berger}, {Brout}, {Challis}, {Drout}, {Finkbeiner}, {Lunnan}, {Kirshner},
  {Sanders}, {Schlafly}, {Smartt}, {Stubbs}, {Tonry}, {Wood-Vasey}, {Foley},
  {Hand}, {Johnson}, {Burgett}, {Chambers}, {Draper}, {Hodapp}, {Kaiser},
  {Kudritzki}, {Magnier}, {Metcalfe}, {Bresolin}, {Gall}, {Kotak}, {McCrum}, \&
  {Smith}}]{Scolnic2018}
{Scolnic}, D.~M., {Jones}, D.~O., {Rest}, A., {et~al.} 2018, Astrophys. J.,
  859, 101, \dodoi{10.3847/1538-4357/aab9bb}

\bibitem[{Shah {et~al.}(2023)Shah, Bhaumik, Mukherjee, \& Pal}]{Shah:2023}
Shah, R., Bhaumik, A., Mukherjee, P., \& Pal, S. 2023, JCAP, 06, 038,
  \dodoi{10.1088/1475-7516/2023/06/038}

\bibitem[{Shah {et~al.}(2024)Shah, Saha, Mukherjee, Garain, \& Pal}]{ladder}
Shah, R., Saha, S., Mukherjee, P., Garain, U., \& Pal, S. 2024, {\texttt{Code:
  LADDER} - Revisiting the Cosmic Distance Ladder with Deep Learning Approaches
  and Exploring its Applications}, 1.0,  Zenodo,
  \dodoi{10.5281/zenodo.11175054}

\bibitem[{Sherwin \& White(2019)}]{Sherwin:2018wbu}
Sherwin, B.~D., \& White, M. 2019, JCAP, 02, 027,
  \dodoi{10.1088/1475-7516/2019/02/027}

\bibitem[{Skidmore {et~al.}(2015)}]{TMT}
Skidmore, W., {et~al.} 2015, Res. Astron. Astrophys., 15, 1945,
  \dodoi{10.1088/1674-4527/15/12/001}

\bibitem[{{Tamanini} {et~al.}(2016)}]{lisa}
{Tamanini}, N., {et~al.} 2016, J. Cosmol. Astropart. Phys., 2016, 002,
  \dodoi{10.1088/1475-7516/2016/04/002}

\bibitem[{Tang {et~al.}(2021)Tang, Li, Lin, \& Liu}]{Tang:2020nmw}
Tang, L., Li, X., Lin, H.-N., \& Liu, L. 2021, Astrophys. J., 907, 121,
  \dodoi{10.3847/1538-4357/abcd92}

\bibitem[{Vagnozzi(2023)}]{Vagnozzi:2023nrq}
Vagnozzi, S. 2023, Universe, 9, 393, \dodoi{10.3390/universe9090393}

\bibitem[{Visser(2005)}]{Visser:2004bf}
Visser, M. 2005, Gen. Rel. Grav., 37, 1541, \dodoi{10.1007/s10714-005-0134-8}

\bibitem[{Wang(2024)}]{Wang:2024rjd}
Wang, D. 2024, arXiv.
\newblock \doarXiv{2404.13833}

\bibitem[{Wang {et~al.}(2020{\natexlab{a}})Wang, Li, \& Xia}]{Wang:2020hmn}
Wang, G.-J., Li, S.-Y., \& Xia, J.-Q. 2020{\natexlab{a}}, Astrophys. J. Suppl.,
  249, 25, \dodoi{10.3847/1538-4365/aba190}

\bibitem[{Wang {et~al.}(2020{\natexlab{b}})Wang, Ma, Li, \& Xia}]{Wang:2019vxv}
Wang, G.-J., Ma, X.-J., Li, S.-Y., \& Xia, J.-Q. 2020{\natexlab{b}}, Astrophys.
  J. Suppl., 246, 13, \dodoi{10.3847/1538-4365/ab620b}

\bibitem[{Wei {et~al.}(2014)Wei, Yan, \& Zhou}]{Wei:2013jya}
Wei, H., Yan, X.-P., \& Zhou, Y.-N. 2014, JCAP, 01, 045,
  \dodoi{10.1088/1475-7516/2014/01/045}

\bibitem[{Xie {et~al.}(2023)Xie, Nong, Zhang, Wang, Li, \& Liang}]{Xie:2023ydk}
Xie, H., Nong, X., Zhang, B., {et~al.} 2023, arXiv.
\newblock \doarXiv{2307.16467}

\bibitem[{{Zhan} \& {Tyson}(2018)}]{lsst2}
{Zhan}, H., \& {Tyson}, J.~A. 2018, Reports on Progress in Physics, 81, 066901,
  \dodoi{10.1088/1361-6633/aab1bd}

\bibitem[{Zhang {et~al.}(2023)Zhang, Xie, Nong, Wang, Xiong, Wu, \&
  Liang}]{Zhang:2023xgr}
Zhang, B., Xie, X., Nong, X., {et~al.} 2023, arXiv.
\newblock \doarXiv{2312.09440}

\bibitem[{Zhang {et~al.}(2017)Zhang, Bengio, Hardt, Recht, \&
  Vinyals}]{augmentation}
Zhang, C., Bengio, S., Hardt, M., Recht, B., \& Vinyals, O. 2017, in
  International Conference on Learning Representations.
\newblock \url{https://openreview.net/forum?id=Sy8gdB9xx}

\bibitem[{Zhang {et~al.}(2024)Zhang, Hu, Jiao, Wang, Xie, Yu, Zhao, \&
  Zhang}]{Zhang:2023ucf}
Zhang, J.-C., Hu, Y., Jiao, K., {et~al.} 2024, Astrophys. J. Suppl., 270, 23,
  \dodoi{10.3847/1538-4365/ad0f1e}

\end{thebibliography}
\bibliographystyle{aasjournal}



\end{document}